\documentclass[conference,a4paper]{IEEEtran}

\usepackage[utf8]{inputenc} 
\usepackage[T1]{fontenc}
\usepackage{url}              % provides \url{...}
\usepackage{cite}             % improves presentation of citations

\usepackage[cmex10]{amsmath}  % Use the [cmex10] option to ensure complicance
\usepackage{mleftright}       % fix to wrong spacing of \left-,
\mleftright                   % \middle- \right-commands 

\usepackage{graphicx}         % provides \includegraphics{...} to
\usepackage{booktabs}         % fixes poor spacing in tables and
\usepackage{xcolor}
\usepackage{amssymb}
\usepackage{epsfig,verbatim}
\usepackage{mathtools}
\usepackage{bbm}
\usepackage{enumitem}

\usepackage{mathtools, cuted}

\long\def\comment#1{}

\newfont{\bbb}{msbm10 scaled 700}

\newfont{\bb}{msbm10 scaled 1100}

\newcommand{\EE}{\mathbb{E}}

\newcommand{\NN}{\mathbb{N}}

\newcommand{\RR}{\mathbb{R}}
\newcommand{\av}{{\bf a}}

\newcommand{\ev}{{\bf e}}

\newcommand{\uv}{{\bf u}}

\newcommand{\vv}{{\bf v}}
\newcommand{\xv}{{\bf x}}
\newcommand{\yv}{{\bf y}}
\newcommand{\zv}{{\bf z}}
\newcommand{\zerov}{{\bf 0}}

\newcommand{\Nm}{{\bf N}}

\newcommand{\Xm}{{\bf X}}
\newcommand{\Ym}{{\bf Y}}
\newcommand{\Zm}{{\bf Z}}

\newcommand{\Ac}{{\cal A}}
\newcommand{\Bc}{{\cal B}}
\newcommand{\Cc}{{\cal C}}

\newcommand{\Hc}{{\cal H}}
\newcommand{\Ic}{{\cal I}}

\newcommand{\Nc}{{\cal N}}

\newcommand{\Pc}{{\cal P}}

\newcommand{\Rc}{{\cal R}}
\newcommand{\Sc}{{\cal S}}

\newcommand{\Uc}{{\cal U}}

\newcommand{\Vc}{{\cal V}}
\newcommand{\Xc}{{\cal X}}
\newcommand{\Yc}{{\cal Y}}

\newcommand{\epsilonv}{\hbox{\boldmath$\epsilon$}}

\newcommand{\var}{{\hbox{Var}}}

\newtheorem{theorem}{Theorem}

\newtheorem{definition}{Definition}

\newtheorem{lemma}{Lemma}

\newtheorem{corollary}{Corollary}

\newtheorem{remark}{Remark}

\newtheorem{proposition}{Proposition}

\newcommand{\argmax}{\operatornamewithlimits{argmax}}
\newcommand{\argmin}{\operatornamewithlimits{argmin}}

\begin{document}

\title{Functional Properties of the Ziv-Zakai bound \\ with Arbitrary Inputs} 

%%%%%%
%\author{%
%  \IEEEauthorblockN{Anonymous Authors}
%  \IEEEauthorblockA{%
%    Please do NOT provide authors' names and affiliations\\
%    in the paper submitted for review, but keep this placeholder.\\
%    ISIT23 follows a \textbf{double-blind reviewing policy}.}
%}

%%%%%% Please only add the author names and affiliations for the FINAL
%%%%%% version of the paper, but NOT for the paper submitted for review!
%
%%%%%
%%%%% Single author, or several authors with same affiliation:
% \author{%
%   \IEEEauthorblockN{Stefan M.~Moser}
%   \IEEEauthorblockA{ETH Zürich\\
%                     8092 Zürich, Switzerland\\
%                     moser@isi.ee.ethz.ch}
%                   }
%
%%%%%
%%%%% Several authors with up to three affiliations:
% \author{%
%   \IEEEauthorblockN{Stefan M.~Moser}
%   \IEEEauthorblockA{ETH Zürich\\
%                     ISI (D-ITET), ETH Zentrum\\
%                     8092 Zürich, Switzerland\\
%                     moser@isi.ee.ethz.ch}
%   \and
%   \IEEEauthorblockN{Albus Dumbledore and Harry Potter}
%   \IEEEauthorblockA{Hogwarts School of Witchcraft and Wizardry\\
%                     Hogwarts Castle\\ 
%                     1714 Hogsmeade, Scotland\\
%                     \{dumbledore, potter\}@hogwarts.edu}
% }
%
%%%%%   
%%%%% Many authors with many affiliations:
 \author{%
   \IEEEauthorblockN{Minoh Jeong\IEEEauthorrefmark{1},
                     Alex Dytso\IEEEauthorrefmark{2},
%                     Stefan M.~Moser\IEEEauthorrefmark{1}\IEEEauthorrefmark{4},
                     and Martina Cardone\IEEEauthorrefmark{1}}
   \IEEEauthorblockA{\IEEEauthorrefmark{1}%
                     University of Minnesota,
                     Minneapolis, MN 55455,
                     \{jeong316, mcardone\}@umn.edu}
   \IEEEauthorblockA{\IEEEauthorrefmark{2}%
                     Qualcomm Flarion Technology, Inc., Bridgewater, NJ 08807,
                     odytso2@gmail.com}
%   \IEEEauthorblockA{\IEEEauthorrefmark{3}%
%                     ETH Zürich, ISI (D-ITET), ETH Zentrum, 
%                     CH-8092 Zürich, Switzerland,
%                     moser@isi.ee.ethz.ch}
%   \IEEEauthorblockA{\IEEEauthorrefmark{4}%
%                     National Yang Ming Chiao Tung University (NYCU), 
%                     Hsinchu, Taiwan,
%                     moser@isi.ee.ethz.ch}
 }

\maketitle

%%%%%
%% Abstract: 
%% If your paper is eligible for the student paper award, please add
%% the comment "THIS PAPER IS ELIGIBLE FOR THE STUDENT PAPER
%% AWARD." as a first line in the abstract. 
%% For the final version of the accepted paper, please do not forget
%% to remove this comment!
%%
\begin{abstract}
This paper explores the Ziv-Zakai bound (ZZB), which is a well-known Bayesian lower bound on the Minimum Mean Squared Error (MMSE).
First, it is shown that the ZZB holds without any assumption on the distribution of the estimand, that is, the estimand does not necessarily need to have a probability density function.
The ZZB is then further analyzed in the high-noise and low-noise regimes and shown to always tensorize.
Finally, the tightness of the ZZB is investigated under several aspects, such as the number of hypotheses and the usefulness of the valley-filling function. In particular, a sufficient and necessary condition for the tightness of the bound with continuous inputs is provided, and it is shown that the bound is never tight for discrete input distributions  with a support set that does not have an accumulation point at zero.
%This paper explores a lower bound of the Minimum Mean Squared Error (MMSE) in estimation problems under the Bayesian perspective. We focus on the {\color{cyan}ZZB}, one of Bayesian lower bound for the MMSE, and generalize the bound applicable to arbitrary inputs not necessarily continuous.
%We further derive the bound from various viewpoints including high and low noise asymptotics and tensorization. We also investigate the tightness of the bound regarding several factors such as the number of hypotheses and the existence of the valley-filling function. In particular, we provide a sufficient and necessary condition for tightness in continuous inputs and demonstrate that the bound is not tight for discrete input distributions.
\end{abstract}

\section{Introduction}
\label{sec:intro}

The mean squared error (MSE) is the {\em de facto} standard metric for many Bayesian estimation problems. Its minimum value, the minimum mean squared error (MMSE) achieved by the conditional expectation, gives an important understanding of how difficult an estimation problem is or on the quality of arbitrary estimators. 
%However, 
Characterizing the MMSE or finding a closed-form of the conditional expectation for arbitrary estimation problems is intractable in general. Therefore, one often needs to rely on lower bounds on the MMSE. A plethora of different bounds has been developed and analyzed. 

We do not attempt to survey the literature, and we only mention a few notable approaches and families of lower bounds. 
A first family of lower bounds relies on the Cauchy-Schwarz inequality, and it was fully explored by Weiss–Weinstein in~\cite{weinstein1988general}. 
We note that the ubiquitous Cram\'er-Rao bound (also known as the Van Trees bound~\cite{van2004detection}) and several others, such as  the Bobrovsky–Zakai bound~\cite{bobrovsky1976lower} and the Bhattacharyya bound~\cite{bhattacharyya1946some} to name a few,
fall under this category. 
A second family of bounds relies on the maximum entropy principle, and it bounds the MMSE with the conditional entropy; see~\cite{Cover:InfoTheory}. 
A  third family of bounds relies on functional inequalities, such as the Poincar\'e inequality~\cite{ISIT2022_Poincare} and the log-Sobolev inequality~\cite{aras2019family}. 
A fourth family of bounds relies on the variational representation of information divergences, such as the $f$-divergence and the $\chi^2$-divergence (resulting in the Bayesian Hammersley-Chapman-Robbins bound); see~\cite{saito2022meta,esposito2021lower,chen2016bayes,xu2016information,polyanskiy2022information}. 
%A fourth family of bounds relies on the variational representation of information divergences, such as the $f$-divergence; see~\cite{saito2022meta,esposito2021lower,chen2016bayes,xu2016information}. 
Finally, a fifth family of bounds, which is under consideration in this work, is known as the Ziv-Zakai bound\footnote{In non-parametric statistics bounds that use hypothesis testing connection are often referred to as Le Cam method~\cite{lecam1973convergence}; see also~\cite{tsybakov2004introduction}.} (ZZB), and it provides a lower bound on the MMSE via a connection to an $\mathsf{M}$-ary hypothesis testing problem~\cite{ZZbound_one_of_the_originals,bell1995performance}.

The most modern and tightest version of the ZZB was presented by Bell et al. in~\cite{bell1995performance}. In particular, one of the main contributions of~\cite{bell1995performance} was to extend the bound to the vector case; prior to this, the bound worked only for the scalar case. One of the key advantages of the ZZB is the fact that it has only one regularity condition:  the parameter under estimation needs to have a probability density function (pdf). Thus, the ZZB is much more general than the Cram\'er-Rao bound, which requires several smoothness assumptions on the pdf of the estimand.  The ZZB is also believed to be one of the tightest bounds in the literature; see~\cite{ZZ_high_noise} for an analysis of the tightness in a high-noise regime. Moreover, despite a somewhat cumbersome expression, the bound is often \emph{not} difficult to evaluate. 
 Because of these advantages, the ZZB has found numerous applications, such as in estimations of the quantum parameter~\cite{tsang2012ziv,PhysRevLett.Giovannetti2012,Gao_2012,PhysRevA.Zhang2014,berry2015quantum,Rubio_2018,PhysRevResearch.Zhang2022,PhysRevLett.Zhuang2022}, time delay~\cite{Mishra2017TDE,9794611}, time of arrival~\cite{Driusso2015TOA,Laas2021TOA,Wang2022TOA,Gifford2022TOA}, position~\cite{Keskin2016VL,Closas2017,gusi2018ziv}, direction of arrival~\cite{Khan2010DOA,Gupta2019DOA,Alexander2019DOA,zhang2022ziv}, and in MIMO radar systems~\cite{Chiriac2010Radar,chiriac2015ziv}.
 
Despite the wide use of the ZZB, several important questions remain unanswered. For example, the ZZB has a free parameter $\mathsf{M}$, which corresponds to the number of auxiliary hypotheses in the bound.   As we increase $\mathsf{M}$, the bound tends to become tighter, but it also becomes more difficult to evaluate. Hence, there is a need to provide practitioners with a rule of thumb on how to choose the parameter $\mathsf{M}$. There are also two versions of the ZZB, one that relies on the so-called \emph{valley-filling function} and another one that does not. The introduction of the valley-filling function makes the bound tighter, but it also complicates the computation of it. Furthermore, it is also not clear how the valley-filling function interacts with the parameter $\mathsf{M}$ and which of these contributes the most to the refinement in the tightness of the ZZB.  Another issue is that the ZZB holds under the condition that the estimand needs to have a pdf, which makes the ZZB unusable in case of discrete or mixed distributions. Mixed distributions play an important role both practically and theoretically in a variety of applications; see, for example,~\cite{wu2012optimal} where mixed distributions are used in compressed sensing. Thus, eliminating the condition would increase the applicability of the bound and make it universal (in the sense that it will not require any regularity conditions). Finally, it is important to provide guidance about the tightness of the ZZB. Some preliminary results on the tightness have been derived in~\cite{bell1995performance} and~\cite{ZZ_high_noise},  and in this work, we make further progress on this front.

\subsection{Contributions and Paper Organization}
We derive and present general properties of the ZZB.
In particular, in Section~\ref{sec:ZZB}, we show that the ZZB holds without any assumption on the distribution of the estimand.  
In the same section, we also characterize the high-noise behavior of the ZZB for a rather general set of noise distributions, as well as a low-noise asymptotics of it for the practically relevant additive Gaussian noise channel model.
Moreover, in Section~\ref{sec:ZZB}, we show the tensorization of the ZZB.
%{\color{cyan}  

In Section~\ref{sec:tightness}, we study the tightness of the bound under several aspects, such as the number of hypotheses and the usefulness of the valley-filling function.
We first focus on a continuous estimand and provide a necessary and sufficient condition under which the ZZB is tight. For a univariate estimand, this improves the result in~\cite{bell1995performance}: it shows that a sufficient condition in~\cite{bell1995performance} is also necessary for the bound to be tight. 
%We also provide an example of input distribution for which $M>2$ does not offer any enhancement of the bound with respect to $M=2$, i.e., a binary hypothesis assumption offers the tightest version of the bound.
% the sufficient condition for tightness of the bound to be a sufficient and necessary condition, and we prove that binary hypothesis setting is enough for input if if has unimodal pdf.}
Then, we move our attention to a discrete estimand.
Surprisingly, for discrete cases, we show that the valley-filling function is a necessary tool; otherwise, the ZZB reduces to zero. We also prove that, even if the valley-filling function is used, the ZZB is never tight for discrete estimand for which the support set does not have an accumulation point at zero.

\subsection{Notation}
Boldface upper case letters $\mathbf{X}$ denote random vectors; 
the boldface lower case letter $\mathbf{x}$ indicates a specific realization of $\mathbf{X}$; 
$X_i$ and $x_i$ (or $(\xv)_i$) denote the $i$th element of $\Xm$ and $\xv$, respectively; 
$\ev_i$ is the $i$th standard basis vector that contains a one in the $i$th entry and a zero in all of the other entries; 
%$X_{i:n}$ denotes the $i$-th order statistics of $\Xm$;
%$\|\Xm\|$ is the norm of $\Xm$ and $\Xm^{\mathsf{T}}$ is the transpose of $\Xm$;
$[n_1: n_2]$ is the set of integers from $n_1$ to $n_2 \geq n_1$;
$\varnothing$ is the empty set;
calligraphic letters $\Xc$ indicate sets/events; 
$\mathsf{d}(\Xc)$ denotes the number of elements of $\xv\in\Xc$;
%$\dim(\Xc)$ denotes the dimension of $\Xc$;
$\left< \xv, \yv \right> = \sum_{i} x_i y_i$ is the inner product between $\xv$ and $\yv$;  
$I_n$ is the identity matrix of dimension $n$.
%Let $\ev_i$ be the $i$-th base vector that contains $1$ in the $i$-th entry and $0$ in the other entries;
%$\{\xv_i\}_{u}^{v}$ for $v \geq u$ is the sequence $\xv_u, \xv_{u+1}, \ldots, \xv_{v-1}, \xv_v$.
For a pair of random vectors $(\Xm,\Ym)$, we let $P_{\Xm}(\cdot)$ denote the distribution of $\Xm$, and $P_{\Ym|\Xm}(\cdot|\xv)$ be the distribution that governs the noisy observation model $\Ym|\Xm=\xv$. 
For a function $f:\RR \to \RR$, the valley-filling function is defined as
\begin{equation}
\label{eq:VF}
	\Vc_t \{ f(t) \}
	 = \sup_{u: u\geq t} f(u).
\end{equation}
%We assume that random vector $\Xm\in\Xc$, distributed according to $P_{\Xm}$, is the parameter that we would seek to estimate and the random vector $\Ym\in\Yc$ is the noisy observation of $\Xm$ through  the channel distribution $P_{\Ym|\Xm}$. 
%The MMSE of estimating $\Xm$ from $\Ym\in\Yc$ is defined as 
%\begin{align}\label{eq:def_mmse}
%	\mathsf{mmse}(\Xm|\Ym)
%	& = \EE[ \| \Xm - \EE[\Xm | \Ym] \|_2^2 ] \nonumber \\
%	& =  \sum_{i=1}^n \EE\left[ \left< \ev_i, \Xm - \EE[\Xm | \Ym] \right>^2 \right],
%\end{align}
%where $\| \cdot\|_2$ is the norm in $\Xc$ defined as similarly to the $\ell^2$-norm in $\RR^n$ (e.g., Frobenius norm if $\Xc = \RR^{m\times n}$), and $\ev_i$'s are the standard basis vectors of $\Xc$ such that $\left<\ev_i,\Xm\right> = X_i,~\forall i$. 
We assume that the random vector $\Xm\in\Xc$
%, distributed according to $P_{\Xm}$, 
is the parameter that needs to be estimated and the random vector $\Ym\in\Yc$ is the noisy observation of $\Xm$ through  the channel distribution $P_{\Ym|\Xm}$. 
The MMSE of estimating $\Xm\in\Xc$ from $\Ym\in\Yc$ is defined as 
\begin{align}\label{eq:def_mmse}
	{\rm mmse}(\Xm|\Ym)
	& =  \sum_{i=1}^{\mathsf{d}(\Xc)} \EE\left[ ( X_i - \EE[X_i | \Ym] )^2 \right].
\end{align}

\section{ZZB with Arbitrary Inputs, Asymptotics, and Tensorization}\label{sec:ZZB}

The presentation of the ZZB requires us to define elements of  an $\mathsf{M}$-ary hypothesis testing setting.  

\begin{definition}
\label{def:MHT}
Let $\mathcal{X}$ be a vector space and $\mathcal{Y}$ be an arbitrary space.  
For random vectors $\Xm\in\Xc$ and $\Ym\in \Yc$, let $P_{\Ym|\Xm}(\cdot|\xv)$ be the conditional probability distribution of $\Ym$ given $\Xm= \xv$. For  an integer $\mathsf{M} \ge 2$ and  $\xv \in \mathcal{X}$, consider the following $\mathsf{M}$-ary hypothesis testing problem, 
\begin{subequations}
\label{eq:hypotest}
\begin{equation}
	\Hc_i: \Ym \sim P_{\Ym|\Xm}(\yv|\xv+\uv_i),  \ i \in [0:\mathsf{M}-1],
\end{equation}
where $\{\uv_i\}_{i=0}^{\mathsf{M}-1}$ is some collection of elements in $\mathcal{X}$, and
\begin{equation}
\Pr(\Hc_i)  = p_i .
\end{equation}
\end{subequations}
For a collection of probabilities $\mathcal{P}=\{ p_i \}_{i=0}^{\mathsf{M}-1}$ and a collection of vectors $\mathcal{U}=\{\uv_i\}_{i=0}^{\mathsf{M}-1} \subset \mathcal{X}$, 
the minimum probability of error for this $\mathsf{M}$-ary hypothesis testing problem (attained with the optimal Maximum a Posteriori (MAP) decision rule~\cite{Kay1998})  is denoted by $P_e \left ({\xv}; \mathcal{P}, \mathcal{U}  \right )$.
% $P_e \left ({\xv;}\mathcal{P}, \mathcal{U} \right )$. 
\end{definition}

\subsection{General ZZB}
\label{sec:General_ZZ}
%
%A general formation of {\color{cyan}ZZB} is given by the following Theorem~\ref{thm:BZZ_Mary_vec_mmse}, which is applicable to any estimation problem with any distribution $\Xm$ and $\Ym$. Here, we present only the {\color{cyan}ZZB} for $\mathsf{mmse}(\Ym|\Xm)$, and a more general version and the proof can be found in Appendix~\ref{app:gen_zzb}. 

%
%Since Theorem~\ref{thm:BZZ_Mary_vec} holds for any estimator $g(\Ym)$ and any {$\av\in \mathcal{X}$,} we can obtain a lower bound on ${\rm mmse}(\Xm|\Ym)$ by taking $g(\Ym) = \EE[\Xm|\Ym]$ and $\av = \ev_i$ in~\eqref{eq:MainThm} and by summing it over $i\in[1:n]$ as stated in the following corollary. 

We present the most general version of the ZZB. This new version holds without any restriction on the joint distribution $P_{\Xm,\Ym}$.  The proof largely depends on the ideas developed in~\cite{bell1995performance}, which assumed that $\Xm$ needs to have a pdf.  By doing a more careful accounting of the terms in the Lebesgue integral, we generalize these results to any probability measure on $ \mathcal{X}$ and prove that the bound holds for any finite-dimensional vector space $\mathcal{X}$.

\begin{theorem}\label{thm:BZZ_Mary_vec_mmse} 
Let $\mathcal{X}$ be a vector space. Then, 
for any integer $\mathsf{M}\geq2$, the ZZB is given by
\begin{equation}
	{\rm mmse}(\Xm|\Ym)
	\geq  \overline{\mathsf{ZZ}}(P_\Xm, P_{\Ym|\Xm},\mathsf{M}),
	%\geq  {\mathsf{LB_{ZZ}^M}}(P_\Xm, P_{\Ym|\Xm}),
\end{equation}
with 
\begin{equation}
	 \overline{\mathsf{ZZ}}(P_\Xm, P_{\Ym|\Xm},\mathsf{M})
	 = \sum_{i=1}^{\mathsf{d}(\Xc)}  \int_0^\infty \frac{t}{2} \Vc_t \left\{ \frac{ \mathsf{h}(t,i,\mathsf{M})}{\mathsf{M}-1} \right\} \ {\rm d} t,  \label{eq:BZZ_Mary_vec_mmse_sub1}
	%& {\mathsf{LB_{ZZ}^M}}(P_\Xm, P_{\Ym|\Xm})
	% = \sum_{i=1}^n  \int_0^\infty \frac{t}{2}  \frac{{\mathsf{h}}(t,\ev_i,M)}{M-1}  {\rm d} t,\label{eq:BZZ_Mary_vec_mmse_sub2}
\end{equation}
where, for $ t >0$ and $ i \in [0:\mathsf{M}-1]$, we have that
 \begin{align}\label{eq:BZZ_Mary_vec_mmse_h}
	\mathsf{h}(t,i,\mathsf{M}) 
	& \! = \!\! \max\limits_{\substack{  \mathcal{U}  \subset \mathcal{X}: \\ \mathcal{U}= \{ \uv_k \}_{k=0}^{\mathsf{M}-1}  \\  (\uv_k)_i= kt,~\forall k}}  \, \int    P_e \left ({\xv}; \mathcal{P}_\mathcal{U}(\xv), \mathcal{U}  \right )  \,    {\rm{d}} \mu_{\mathcal{U}} (\xv).   
\end{align}
To define $h(t,i,\mathsf{M})$, we let:
\begin{itemize}[leftmargin=*]
\item for a set $\mathcal{U}= \{ \uv_k \}_{k=0}^{\mathsf{M}-1} \subset \mathcal{X} $, the measure $\mu_\Uc$ is given by
%we define the following measure,
\begin{equation}
\mu_{\mathcal{U}} =\sum_{{k=0}}^{\mathsf{M}-1}  P_{\Xm-\uv_{k} }; 
\end{equation} 
\item   for $\xv \in \mathcal{X}$ and $\mathcal{U} \subset \mathcal{X}$, we let\footnote{By construction, the Radon-Nikodym derivative is always well defined.}
 \begin{equation}\label{eq:def_P_U}
 \mathcal{P}_\mathcal{U}(\xv) \!=\! \left \{ p_i\!:\! p_i \!=\!  \frac{  {\rm{d}} P_{\Xm-\uv_{i} } }{   {\rm{d} \mu_{\mathcal{U}}}   } (\xv),  \! \uv_i \in \mathcal{U}\!=\! \{ \uv_k \}_{k=0}^{\mathsf{M}-1}
 \right \}.
 \end{equation} 
% note that by construction, the Radon-Nikodym derivative is always well defined.
% \item  we let 
\end{itemize} 

 %are the prior probabilities for the hypothesis testing in Definition~\ref{def:MHT}.
%\begin{align}\label{eq:BZZ_Mary_vec_mmse}
%	mmse(\Xm|\Ym)
%	\geq \sum_{i=1}^n \int_0^\infty \frac{t}{2} G\left( \frac{1}{M-1} \max_{\substack{ \uv_0,...,\uv_{M-1} \\ \ev_i^{\mathsf{T}} \uv_k = kt,~ \forall k}} \int \left( \sum_{j=0}^{M-1} f_\Xm(\xv+\uv_j) \right) P_e^{\rm opt}(\xv,\uv_0,...,\uv_{M-1}) {\rm d} \xv \right) {\rm d} t,
%\end{align}
%$\Vc\{\cdot\}$ is the valley-filling function, and $P_e(\eta,\{\xv + \uv_i\}_0^{M-1})$ is the minimum error probability for the $M$-ary hypothesis testing problem defined in~\eqref{eq:hypotest}.
\end{theorem}
\begin{IEEEproof}
The proof is provided in Appendix~\ref{app:gen_zzb}.
\end{IEEEproof}
%The result in Theorem~\ref{thm:BZZ_Mary_vec_mmse} was already known for the case when $\mathcal{X} = \mathbb{R}^n$ and $\Xm$ is a continuous random vector~\cite[eq.~(3.85), eq.~(3.86)]{bell1995performance}. In Theorem~\ref{thm:BZZ_Mary_vec_mmse}, we have generalized the Ziv-Zakai lower bound for any probability measure on $\Xm \in \mathcal{X}$ and proved that it holds for any vector space $\mathcal{X}$.
\begin{remark}
The ZZB in Theorem~\ref{thm:BZZ_Mary_vec_mmse} holds for any vector space $\Xc$, whenever the ${\rm mmse}(\Xm|\Ym)$ for $\Xm\in\Xc$ is properly defined as in~\eqref{eq:def_mmse}.
For instance, $\Xm\in\Xc$ can follow a mixture distribution on an arbitrary finite-dimensional vector space, such as a vector, matrix, or tensor.
In addition, it is worth noting that removing the valley-filling function in~\eqref{eq:BZZ_Mary_vec_mmse_sub1} further loosens but also simplifies the bound. The version without the valley-filling function is defined as,
\begin{equation}
	{\mathsf{ZZ}}(P_\Xm, P_{\Ym|\Xm},\mathsf{M})
	= \sum_{i=1}^{\mathsf{d}(\Xc)}  \int_0^\infty \frac{t}{2}  \frac{{\mathsf{h}}(t,i,\mathsf{M})}{\mathsf{M}-1} \ {\rm d} t. \label{eq:BZZ_Mary_vec_mmse_sub2}
\end{equation}
\end{remark}
Although the ZZB in Theorem~\ref{thm:BZZ_Mary_vec_mmse} has a rather complex expression, one can evaluate it easily if the exact (or an approximate) expression for $ P_e \left ({\xv}; \mathcal{P}_\mathcal{U}(\xv), \mathcal{U}  \right )$ is available. 
However, computing the ZZB in Theorem~\ref{thm:BZZ_Mary_vec_mmse} for every  $P_{\Ym|\Xm}$ might be difficult. To provide guidance on its tightness, one can resort to understanding if it is tight in asymptotic regimes.
%the low and high noise regimes. 
The high and low noise asymptotics of the ZZB are characterized next.

\subsection{High-Noise Asymptotics}
\label{sec:High_Noise_Assymptotics}
Here, we study the ZZB in Theorem~\ref{thm:BZZ_Mary_vec_mmse}  in the high-noise regime. 
In particular, we characterize the asymptotics under the following assumptions (rather natural as discussed in~\cite{ZZ_high_noise}):
%In particular, we restrict ourselves to noise distributions for which the following (rather natural as discussed in~\cite{ZZ_high_noise}) assumptions hold:
%In order to express such regime in limit expression, we adopt the following assumptions:
\begin{itemize}
	\item {\bf A1:} $P_{\Ym|\Xm}$ can be parameterized by $\eta\geq0$, which is referred to as the {\em noise level}, i.e., $P_{\Ym|\Xm}(\yv|\xv;\eta)$ for all $(\xv,\yv)\in(\Xc,\Yc)$. 
	In order to highlight the dependence on~$\eta$, we let $P_e \left ({\xv}; \mathcal{P}_\mathcal{U}(\xv), \mathcal{U}  \right ) = P_e \left (\eta,{\xv}; \mathcal{P}_\mathcal{U}(\xv), \mathcal{U}  \right ) $ denote the optimal error probability in Definition~\ref{def:MHT}. \label{item:A1}
	\item {\bf A2:} $P_e \left (\eta,{\xv}; \mathcal{P}_\mathcal{U}(\xv), \mathcal{U}  \right ) $ is non-decreasing in $\eta$. \label{item:A2}
	\item {\bf A3:} $\lim\limits_{\eta\to\infty} P_e \left (\eta, {\xv}; \mathcal{P}_\mathcal{U}(\xv), \mathcal{U}  \right )  = 1 -  \max\limits_{p\in\Pc_\Uc(\xv)} p $. \label{item:A3}
\end{itemize}
%{\color{blue} A.D. We need to say that the high assumptotics section is a generalization of our Singnal Processing Paper  technique}
Next, we define the high-noise asymptotics of the ZZB.
\begin{definition}\label{def:high_noise}
The high-noise asymptotics of the ZZB in Theorem~\ref{thm:BZZ_Mary_vec_mmse} is defined as
\begin{align}
	\overline{\mathsf{V}}(P_\Xm,\mathsf{M}) = \lim_{\eta\to\infty} \overline{\mathsf{ZZ}}(P_\Xm, P_{\Ym|\Xm},\mathsf{M}) , \\
	{\mathsf V}(P_\Xm,\mathsf{M}) = \lim_{\eta\to\infty} {\mathsf{ZZ}}(P_\Xm, P_{\Ym|\Xm},\mathsf{M})  . 
\end{align}
%Note that the bound does not have a dependency on $P_{\Ym|\Xm}$ in high-noise asymptotic regime.
\end{definition}
%The high-noise asymptotics of the {\color{cyan}ZZB}, whose proof is in Appendix~\ref{app:high_noise}, is given next.  
%
The next theorem characterizes the asymptotics in Definition~\ref{def:high_noise}.
\begin{theorem}\label{thm:zzb_sig_inf_M}
For any $P_\Xm$ and any $\mathsf{M}\ge 2$, we have that
%\begin{subequations}
\begin{align*}
	& \overline{\mathsf V}(P_\Xm,\mathsf{M})
	=   \sum_{i=1}^{\mathsf{d}(\Xc)}  \int_0^\infty \frac{t}{2} \Vc_t\left\{ \frac{ \mathsf{M} - \mathsf{H}(P_\Xm,t,i,\mathsf{M})  }{\mathsf{M}-1} \right\} \ {\rm d} t, \\
	& {\mathsf V}(P_\Xm,\mathsf{M})
	=   \sum_{i=1}^{\mathsf{d}(\Xc)} \int_0^\infty \frac{t}{2}  \frac{ \mathsf{M} - \mathsf{H}(P_\Xm,t,i,\mathsf{M})  }{\mathsf{M}-1} \ {\rm d} t,
\end{align*}
%\end{subequations}
where
\begin{align*}
%\label{eq:h_high_noise}
	\mathsf{H}(P_\Xm,t,i,  \mathsf{M})
	& =  \min\limits_{\substack{  \mathcal{U}  \subset \mathcal{X}: \\ \mathcal{U}= \{ \uv_k \}_{k=0}^{\mathsf{M}-1}  \\   (\uv_k)_i =kt,~\forall k }}  \int  \max_{j\in[0:\mathsf{M}-1]}  {\rm{d}} P_{\Xm-\uv_{j} }(\xv)  .
\end{align*}
%\begin{align}
%	\lim_{\eta\to\infty} \overline{\mathsf{LB_{ZZ}}}(\Xm,\eta)
%	& =  \sum_{i=1}^n \int_0^\infty  \frac{t}{2} G\left( \frac{1}{M-1} \max_{\substack{ \uv_0,...,\uv_{M-1} \\ \ev_i^{\mathsf{T}} \uv_k = kt,~ \forall k}}   \left( M - \int  \max_{i\in[0:M-1]} f_\Xm(\xv+\uv_i)  {\rm d} \xv \right)  \right) {\rm d} t .
%\end{align}
\end{theorem}
\begin{IEEEproof}
The proof is provided in Appendix~\ref{app:high_noise}.
\end{IEEEproof}
 It is worth noting that, in the high-noise regime, we have that
 % asymptotics of MMSE corresponds to the variance of $\Xm$, which can be expressed as
\begin{equation}
	\lim_{\eta \to \infty} {\rm mmse}(\Xm | \Ym) 
%	& = \sum_{i=1}^{\dim(\Xc)} \EE \left[( X_i - \EE[ X_i ] )^2\right] \nonumber \\
	= \sum_{i=1}^{\mathsf{d}(\Xc)}\var(X_i),
\end{equation}
and hence, the ZZB would be tight in this regime if either one of the quantities ${\mathsf V}(P_\Xm,\mathsf{M})$ or $ \overline{\mathsf V}(P_\Xm,\mathsf{M})$ is equal to the variance.
In~\cite{ZZ_high_noise}, the authors have provided a few examples for the univariate case for which the ZZB is tight and not tight in the high-noise regime.

We now conclude the analysis of the high-noise regime by showing that Theorem~\ref{thm:zzb_sig_inf_M} can be helpful in understanding the effect of $\mathsf{M}$. In particular, we provide an example of input distribution for which $\mathsf{M} > 2$ does not offer any enhancement of the bound with respect to $\mathsf{M} = 2$.
%, i.e., a binary hypothesis assumption offers the tightest version of the bound.

\noindent {\bf{Example.}} Let $f_{X}$ be the pdf of $X$ and assume that $f_{X}$ is unimodal. Then, 
\begin{align*}
	& \overline{\mathsf V}(P_X,  \mathsf{M})
	= \overline{\mathsf V}(P_X,2) , \text{ and }
%\nonumber \\
%	& =   \int_0^\infty \frac{t}{2} \Vc_t\left\{ 1 - F_{X}(a(t)+t) + F_{X}(a(t)) \right\} {\rm d} t, \\
	 {\mathsf V}(P_X,\mathsf{M})
	= {\mathsf V}(P_X,2),
%	& =   \int_0^\infty \frac{t}{2}  (1 - F_{X}(a(t)+t) + F_{X}(a(t))  ) {\rm d} t.
\end{align*}
which can be proved by leveraging Lemma~\ref{lem:unimodal_pdf} in Appendix~\ref{app:Lemmas}.

\subsection{Low-Noise Asymptotics}
\label{sec:Low_Noise_Assymptotics}
Here, we study the ZZB in the low-noise regime, i.e., when $\eta \to 0$. 
We start by noting that $\lim_{\eta\to 0} {\rm mmse}(\Xm | \Ym) = 0$ and hence, our focus is on characterizing the rate of convergence. However, since such a convergence rate highly depends on the noise distribution, we here focus on the practically relevant additive Gaussian noise channel.
%As the MMSE tends to be zero in the low-noise regime (i.e., $\lim_{\eta\to 0} {\rm mmse}(\Xm | \Ym) = 0$), we instead characterize the rate of convergence of Ziv-Zakai lower bound under the standard Gaussian noise. 
In particular, we consider
\begin{align}\label{eq:AWGN}
	\Ym = \Xm +\Nm, \text{ where } \Nm\sim\Nc(\zerov, {\eta I_d}),
\end{align}
where $d = {\mathsf{d}(\Xc)}$ and $\eta$ is the noise level defined in Section~\ref{sec:High_Noise_Assymptotics}.
%{\color{cyan} Can we use $\Nc(\zerov, \sigma^2 I_d)$? This is a standard notation for Gaussian distribution. }
The next theorem proves the low-noise asymptotics of the ZZB for the channel model in~\eqref{eq:AWGN}.
\begin{theorem}\label{thm:low_noise}
Consider any $P_\Xm$  such that
%be a distribution function of $\Xm\in\Xc$, and it can be written as
\begin{align}
\label{eq:MixDistr}
	P_{\Xm} = \alpha P_{\Xm_C} + (1-\alpha)P_{\Xm_D},~0\leq \alpha \leq 1,
\end{align}
where $P_{\Xm_C}$ is an absolutely continuous distribution with respect to an $d$-dimensional Lebesgue measure, and $P_{\Xm_D}$ is a purely discontinuous distribution.
Then, for the channel model in~\eqref{eq:AWGN}, it holds that
\begin{align}
	\lim_{\eta \to0} \frac{ {\mathsf{ZZ}}(P_\Xm, P_{\Ym|\Xm},2 )}{\eta} 
	% {\mathsf{LB_{ZZ}^M}}(P_\Xm, P_{\Ym|\Xm})
	& = \alpha \mathsf{d}(\Xc).
%\dim(\Xc).
\end{align}
\end{theorem}
\begin{IEEEproof}
The proof is provided in Appendix~\ref{app:low_noise_proof}.
\end{IEEEproof}

%\begin{IEEEproof}
%The proof of Theorem~\ref{thm:low_noise} is in Appendix~\ref{app:low_noise_proof}.
%\end{IEEEproof}

\begin{remark}
In~\cite{David2016,MMSEdim}, it was shown that under the channel model in~\eqref{eq:AWGN} and $P_\Xm$ as in~\eqref{eq:MixDistr}, it holds that
%Theorem~\ref{thm:low_noise},}
%the low-noise asymptotics of ${\rm mmse}(\Xm|\Ym)$ in AWGN channel~\eqref{eq:AWGN} is given by
\begin{align}
	\lim_{\eta \to0} \frac{{\rm mmse}(\Xm|\Ym)}{\eta} = \alpha \mathsf{d}(\Xc).
%\dim(\Xc).
\end{align}
Thus,  Theorem~\ref{thm:low_noise} demonstrates that the ZZB is tight in the low-noise regime under the channel model in~\eqref{eq:AWGN}. Moreover, in the low-noise regime, the ZZB can be used in its simplest form in~\eqref{eq:BZZ_Mary_vec_mmse_sub2} with $\mathsf{M}=2$. 
Furthermore, we note that while for continuous distributions, there are many bounds that are tight in the low-noise regime (e.g., Cram\'er-Rao), we are not aware of any lower bounds that are tight in the low-noise regime for mixed distributions other than the ZZB. We note that mixed distributions are important in compressed sensing applications (see, for example, \cite{wu2012optimal}). 
\end{remark}

\subsection{Tensorization}
\label{sec:tensorization}
Due to the cumbersome expression of the ZZB, particularly the presence of an inner layer of optimization in~\eqref{eq:BZZ_Mary_vec_mmse_h},  it is important to understand whether the bound admits any simplification. The next proposition, which also concludes this section, indeed shows that the ZZB tensorizes. 
%tensorizes. 
%This property is indeed shown by the next proposition, the proof of which can be found in Appendix~\ref{app:sec:tensorization}. 
\begin{proposition}\label{prop:tensorization}
Whenever $P_\Xm = \prod_{i=1}^d P_{X_i}$ and $P_{\Ym|\Xm} = \prod_{i=1}^d P_{Y_i|X_i}$ with $d = \mathsf{d}(\Xc)$, it holds that
\begin{subequations}
\begin{align}
	\overline{\mathsf{ZZ}}(P_{\Xm}, P_{\Ym|\Xm},\mathsf{M})
	 =  \sum_{i=1}^{d} \overline{\mathsf{ZZ}}(P_{X_i}, P_{Y_i| X_i},\mathsf{M}), \\
	 {\mathsf{ZZ}}(P_{\Xm}, P_{\Ym|\Xm},\mathsf{M})
	 =  \sum_{i=1}^{d} {\mathsf{ZZ}}(P_{X_i}, P_{Y_i| X_i},\mathsf{M}) .
\end{align}
\end{subequations}
\end{proposition}
\begin{IEEEproof}
The proof is provided in Appendix~\ref{app:sec:tensorization}.
\end{IEEEproof}
%{\color{red}MC: Can we add some comment here on why the above result is useful? Also, can we add a paragraph saying that this result is different from the CR that does not tensorize?}
%{\color{cyan} I think that the sentence about the inner layer of optimization before the Proposition 1 shows its usefulness. }

%{\color{blue} I removed everything else that was here as it was trivial.} 

%Note that if the property of tensorization is applicable, one can avoid the maximization in $\mathsf{h}(t,\ev_i,M)$ in~\eqref{eq:BZZ_Mary_vec_mmse_h} since, for an univariate $X$ and $Y$, it becomes that
%\begin{align}
%	\mathsf{h}(t,\ev_i,M)  
%	& \! = \!\! \int  \!  \sum_{{k=0}}^{M-1} {\rm{d}} P_{X-kt }(x)  P_e \left(x; \{ f_j  \}_j, \{ kj\}_j, M\right) .
%\end{align}
%
%
%As shown in Proposition~\ref{prop:tensorization}, the product probability measure together with independent channel implies the tensorization of {\color{cyan}ZZB}. The high-noise regimes, which only depends on the prior distribution rather than channel, indeed needs a product probability measure of prior distribution in order for the high-noise asymptotics to be tensorized without the independence condition on the channel. Formally, the tensorization of the high-noise asymptotics is stated in the following Corollary.
%
%\begin{corollary}\label{cor:ZZB_high_asymptotic_tensor}
%If $P_{\Xm} = \prod_i P_{X_i}$, it holds that
%\begin{align}
%	& \overline{\mathsf V}(P_\Xm,M) = \sum_i \overline{\mathsf V}(P_{X_i},M) \\
%	& {\mathsf V}(P_\Xm,M) = \sum_i {\mathsf V}(P_{X_i},M) .
%\end{align}
%\end{corollary}

\section{Tightness Analysis}\label{sec:tightness}
In this section, we study the tightness of the ZZB in Theorem~\ref{thm:BZZ_Mary_vec_mmse}  for various distribution settings of $\Xm$, namely continuous distributions and discrete distributions. 

\subsection{Continuous Distribution}
\label{sec:Tigtness_in_general}
We assume that $P_\Xm$ and $P_{\Ym|\Xm}$ are distributions with pdfs $f_\Xm$ and $f_{\Ym|\Xm}$, respectively.
%continuous distributions. Throughout this section, we respectively denote the pdf of $\Xm$ and $\Ym|\Xm$ by $f_\Xm$ and $f_{\Ym|\Xm}$.
In the following proposition\footnote{We remind the reader that, in general,  the $\argmax$ function outputs a set. }, we present a sufficient and necessary condition on $f_{\Xm|\Ym}$ that results in a tight ZZB on the ${\rm mmse}(\Xm|\Ym)$.

\begin{proposition}\label{prop:mmse=ZZB}
For $\Xm\in\Xc$ and $\Ym\in\Yc$, it holds that
\begin{align}\label{eq:mmse=ZZB}
	{\rm mmse}(\Xm|\Ym) = {\mathsf{ZZ}}(P_\Xm, P_{\Ym|\Xm},\mathsf{M}),
\end{align}
if and only if for all $t>0$ there exist $\Uc_i = \{\uv_{i,k}\}_{k=0}^{\mathsf{M}-1} \subset \Xc,~\forall i\in[1: \mathsf{d}(\Xc)]$, each of which satisfies $( \uv_{i,k})_i= kt,~k\in[0:\mathsf{M}-1]$ and for all $(\xv,\yv)\in\Xc\times\Yc$,
\begin{align}\label{eq:suff_necess_cond}
	&  \argmax_{k\in[0:\mathsf{M}-1]} f_{\Xm|\Ym}(\xv+\uv_{i,k} |  \yv) \nonumber \\
	& \cap \argmin_{k\in[0:\mathsf{M}-1]} |\EE[X_i|\Ym = \yv] - x_i - kt |  \neq \varnothing.
\end{align}
\end{proposition}
%{\color{magenta}
%\begin{align*}
%\mathcal{S}_{\mathsf{MAP}}(\{\xv_i\}_{i=0}^{M-1},\yv) &= \argmax_{k\in[0:M-1]} P_{\Xm|\Ym}(\xv_k | \Ym = \yv)
%\\ \mathcal{S}_{\mathsf{MMSE}}(\xv,\yv) & = \argmin_{k\in[0:\mathsf{M}-1]} |\EE[X_i|\Ym = \yv] - x_i - kt | 
%%  $\mathcal{S}_{\mathsf{MAP}}(\xv+u_{i,k},\yv) \cap \mathcal{S}_{\mathsf{MMSE}}(\xv,\yv) \neq \varnothing
%\end{align*}
%}

\begin{IEEEproof}
We start by noting that the ZZB in Theorem~\ref{thm:BZZ_Mary_vec_mmse}  was derived by individually lower bounding each of the $d= \mathsf{d}(\Xc)$ elements that contribute to ${\rm mmse}(\Xm|\Ym)$ (see~\eqref{eq:def_mmse}).  Thus,~\eqref{eq:mmse=ZZB} holds if and only if the $i$th element that contributes to ${\rm mmse}(\Xm|\Ym)$ is equal to the $i$th element that contributes to ${\mathsf{ZZ}}(P_\Xm, P_{\Ym|\Xm},\mathsf{M})$, where $i\in[1: d]$.

%for each of the $d=\dim(\Xc)$ terms that contribute to ${\rm mmse}(\Xm|\Ym)$ there is an equal term that contributes to ${\mathsf{ZZ}}(P_\Xm, P_{\Ym|\Xm},\mathsf{M})$.
% (since all of these terms are non-negative).
Now, consider the $i$th element that contributes to ${\mathsf{ZZ}}(P_\Xm, P_{\Ym|\Xm},\mathsf{M})$ in~\eqref{eq:BZZ_Mary_vec_mmse_sub2}.
This term was derived by bounding only the probability of error in ${\mathsf{h}}(t,i,\mathsf{M})$ (see~\eqref{eq:Before_valley} in Appendix~\ref{app:gen_zzb}) as
\begin{align}
\label{eq:ProbIneq}
%\Pr\left( |\left< \av, \epsilonv \right>| \geq \frac{t}{2} \right)
   P_e^{\phi} \left ({\xv}; \mathcal{P}_\mathcal{U}(\xv), \mathcal{U} \right )  
& \ge  P_e \left ({\xv}; \mathcal{P}_\mathcal{U}(\xv), \mathcal{U} \right ),
\end{align}
% which is $ \int_0^\infty \frac{t}{2}  \frac{{\mathsf{h}}(t,i,\mathsf{M})}{\mathsf{M}-1}  {\rm d} t$.
%${\mathsf{ZZ}}(P_\Xm, P_{\Ym|\Xm},\mathsf{M})$ was derived 
where $P_e^{\phi} \left ({\xv}; \mathcal{P}_\mathcal{U}(\xv), \mathcal{U} \right )$ is the error probability incurred by the sub-optimal decision rule $\phi(\Ym)$ given by
\begin{align}\label{eq:phi_sub_opt}
	\phi(\Ym) = \Hc_j, \text{ where } j \in \!\! \argmin_{k\in[0:\mathsf{M}-1]} \! |\EE[X_i |\Ym] - x_i - kt|,
%	\begin{cases}
%		\xv + \uv_0  & \text{ if } \EE[X_i | \Ym] - x_i  <  \frac{t}{2} \\
%		\xv + \uv_k & \text{ if } \left(k - \frac{1}{2}\right)t < \EE[X_i | \Ym] - x_i   < \left( k + \frac{1}{2}\right)t \\
%		\xv + \uv_{M-1} & \text{ if } \left( M-\frac{3}{2}\right)t < \EE[X_i | \Ym] - x_i  
%	\end{cases}.
\end{align}
and $P_e\left ({\xv}; \mathcal{P}_\mathcal{U}(\xv), \mathcal{U} \right )$ is the minimum error probability, i.e.,  incurred by the optimal MAP decision rule $\phi^\star(\Ym)$, that is
\begin{align}\label{eq:phi_opt}
	\phi^\star(\Ym)
	& = 
	\Hc_j,  \text{ where } j \in \argmax_{k\in[0:\mathsf{M}-1]} f_{\Xm|\Ym}(\xv+\uv_k | \Ym ).
\end{align}
In other words,  $P_e^{\phi} \left ({\xv}; \mathcal{P}_\mathcal{U}(\xv), \mathcal{U} \right )$ leads to ${\mathsf{ZZ}}(P_\Xm, P_{\Ym|\Xm},\mathsf{M})$ and $P_e \left ({\xv}; \mathcal{P}_\mathcal{U}(\xv), \mathcal{U} \right )$ leads to ${\rm mmse}(\Xm|\Ym)$.
Thus, the equality in~\eqref{eq:mmse=ZZB} holds if and only if~\eqref{eq:ProbIneq} holds with equality for all $i \in [1:d]$ and hence, if and only if $\phi=\phi^\star$ for all $i \in [1:d]$.

%Hence, the equality~\eqref{eq:mmse=ZZB} implies that the decision rule~\eqref{eq:phi_sub_opt} is indeed MAP decision rule $\phi^\star$ that is given by
%\begin{align}
%	\phi^{\rm MAP}(\Ym)
%	& = 
%	\xv + \uv_i,  \text{ where } i = \max_{w\in[0:M-1]} \Pr(\Hc_w | \Ym) .
%\end{align}
%Since we can write
%\begin{align}
%	\Pr(\Hc_w | \Ym)
%	& = \frac{\Pr(\Ym | \Hc_w) \Pr(\Hc_w)}{ P_{\Ym}(\Ym)} \nonumber \\
%	& = \frac{ P_{\Ym|\Xm}(\Ym | \Xm = \xv+\uv_w) }{ P_{\Ym}(\Ym)} \frac{P_{\Xm - \uv_w}(\xv)}{\sum_{{k=0}}^{M-1} {\rm{d}} P_{\Xm-\uv_{k} }(\xv)} ,
%\end{align} 
%the MAP decision rule in our case is equivalent to

Now, since $\mathsf{h}(t,i,\mathsf{M})$ is maximized over $\Uc = \{\uv_k\}_{k=0}^{\mathsf{M}-1}\subset \Xc$, for all $t>0$, we need at least one set $\Uc$ (recall that $t$ is a constraint for $\Uc$) such that $\phi(\yv)=\phi^\star(\yv)$ for all $\yv\in\Yc$. 
It therefore follows that a sufficient and necessary condition for the tightness of the $i$th element that contributes to ${\mathsf{ZZ}}(P_\Xm, P_{\Ym|\Xm},\mathsf{M})$
% (only the $i^{\rm th}$ term of ${\mathsf{ZZ}}(P_\Xm, P_{\Ym|\Xm},\mathsf{M})$) 
is given by that for all $t>0$, there exist $\Uc_i = \{\uv_{i,k}\}_{k=0}^{\mathsf{M}-1}$ such that $(\uv_{i,k} )_i = kt,~k\in[0:\mathsf{M}-1]$ and for all $ (\xv,\yv)\in\Xc\times \Yc$,
\begin{align}
	&  \argmax_{k\in[0:\mathsf{M}-1]} f_{\Xm|\Ym}(\xv+\uv_{i,k} | \Ym = \yv) \nonumber \\
	& \cap  \argmin_{k\in[0:\mathsf{M}-1]} |\EE[X_i|\Ym = \yv] - x_i - kt | \neq \varnothing.
\end{align}
Hence,~\eqref{eq:mmse=ZZB} holds if and only if there exist such $\Uc_i$ for all $i \in[1:d]$. This concludes the proof of Proposition~\ref{prop:mmse=ZZB}.
\end{IEEEproof}

\begin{remark}
In words, the condition in~\eqref{eq:suff_necess_cond} implies that the $i$th element of the MAP estimate over $\Sc = \{\xv + \uv_{i,k}\}_{k=0}^{\mathsf{M}-1}$ should be close to the MMSE estimate of $X_i$ given $\Ym$.
%In words, the condition in~\eqref{eq:suff_necess_cond} implies that the MAP estimate over the set $\Sc = \{\xv + \uv_{i,k}\}_{k=0}^{\mathsf{M}-1}$ should be the closest point to the MMSE estimate of $X_i$ given $\Ym$ over $\Sc$.
\end{remark}

By leveraging Proposition~\ref{prop:mmse=ZZB}, we now strengthen a result in~\cite{bell1995performance}, namely we prove that a condition that was shown in~\cite{bell1995performance} to be sufficient for the tightness of the ZZB is indeed also necessary.
%the sufficient condition~\cite{bell1995performance} for tight {\color{cyan}ZZB} by showing that it is also a necessary condition for tightness of {\color{cyan}ZZB}.
\begin{corollary}\label{cor:unimodal_symmetric}
Let $X\in\RR$ be a continuous random variable. Then, it holds that
\begin{equation}\label{eq:unimodal_symmetric}
	{\rm mmse}(X|\Ym) = {\mathsf{ZZ}}(P_X, P_{\Ym|X},\mathsf{M}),
\end{equation}
if and only if, for all $\yv \in \Yc$,  the pdf $f_{X|\Ym}(x|\yv)$  is unimodal\footnote{When a pdf has multiple local maxima, it is common to refer to all of them as modes of the distribution.} and symmetric with respect to its mode.
\end{corollary}
\begin{IEEEproof}
The proof is provided in Appendix~\ref{app:CorIff}.
\end{IEEEproof}

Finally, we note that sufficient and necessary conditions for the exact equality of the bound with the valley-filling function are unknown and constitute an interesting future direction.

\subsection{Discrete Distribution}
\label{sec:Discrete_Distributions}

%As the {\color{cyan}ZZB} in Theorem~\ref{thm:BZZ_Mary_vec_mmse} holds for discrete $\Xm$, we should analyze how tight the bound is for discrete $\Xm$, whereas its tightness for continuous random variables has been studied in numerous literature~\cite{tightness for countinuous}. An important component of the {\color{cyan}ZZB} is the valley-filling function (i.e., $\Vc_t\{\cdot\}$), and we strengthen its significance in the following proposition by showing that it is indeed indispensable to the bound for discrete $\Xm$, otherwise the bound becomes trivial.

The ZZB has been generally studied under the assumption that the distribution of $\Xm$ has a pdf~\cite{bell1995performance}. The case of discrete random variables is typically ignored and  often erroneously assumed to be as trivial as replacing the pdf with the corresponding probability mass function (pmf). The goal of this part of the paper is to  show that care needs to be taken when dealing with discrete distributions, and this,  in part, is why Theorem~\ref{thm:BZZ_Mary_vec_mmse} was presented in a measure-theoretic way.

We now present the following  series of results. First, we prove that the ZZB without the valley-filling function is always equal to zero. Second, we present a simple example that shows that the bound with the valley-filling function is not equal to zero. These two results imply that for discrete inputs, the valley-filling function is an indispensable component of the ZZB. Third,  the final result in this series shows that even with the valley-filling function, the ZZB is  strictly sub-optimal for discrete inputs.  
\begin{proposition}\label{prop:discrete_zero_zzb}
Suppose that  $\Xm$ is discrete. Then, 
%for any $\mathsf{M} \geq2$,
\begin{equation}
	{\mathsf{ZZ}}(P_\Xm, P_{\Ym|\Xm},\mathsf{M})
	 = 0.
\end{equation}
\end{proposition}
\begin{IEEEproof}
Let $p_{\Xm}$ denote the pmf of $\Xm \in \Xc$ with $d= \mathsf{d}(\Xc)$.
For a discrete $\Xm$,  from Theorem~\ref{thm:BZZ_Mary_vec_mmse},
% the Ziv-Zakai lower bound without the valley-filling function is given by 
%\begin{align}\label{eq:discrete_zero_proof1}
%	{\mathsf{ZZ}}(P_\Xm, P_{\Ym|\Xm},\mathsf{M})
%	 = \sum_{i=1}^n  \int_0^\infty \frac{t}{2}  \frac{ {\mathsf{h}}(t,i,{\mathsf{M}}) }{\mathsf{M}-1}  {\rm d} t,
%\end{align}
%To ensure~\eqref{eq:discrete_zero_proof1} being $0$, it suffices to show that ${\mathsf{h}}(t,\ev_i,M) = 0$ almost everywhere in $t\geq0$, or equivalently we need the following sufficient condition that 
%\begin{align}\label{eq:discrete_zero_proof2}
%	\Ac = \{t\geq0 : \mathsf{h}(t,\ev_i,M) >0 \}
%\end{align}
%is countable. 
%For a discrete $\Xm$ we can write $\mathsf{h}(t,\ev_i,M) $ as
%where for a discrete $\Xm$ 
we have that
\begin{equation}\label{eq:discrete_zero_proof3}
	\mathsf{h}(t,i,\mathsf{M})  
	 \!=\!\!\! \max\limits_{\substack{ \uv_k \in \mathcal{X},\forall k  \\  (\uv_k )_i = kt}}  \sum_{\xv\in\overline{\Xc} }    \sum_{{k=0}}^{\mathsf{M}-1} \! p_{\Xm-\uv_{k} } \!(\xv)  P_e \left(\xv;\Pc_\Uc(\xv),\Uc \right),
\end{equation}
where $\overline{\Xc} = \bigcup_{i=1}^d\{\Xc-\uv_i\}$.   We will demonstrate that $   P_e \left(\xv;\Pc_\Uc(\xv),\Uc \right)$ in~\eqref{eq:discrete_zero_proof3} equal to zero almost surely for $t \geq 0$, which, via \eqref{eq:BZZ_Mary_vec_mmse_sub2} and~\eqref{eq:discrete_zero_proof3},  implies that ${\mathsf{ZZ}}(P_\Xm, P_{\Ym|\Xm},\mathsf{M})= 0$.

%From the above, we have that a sufficient condition for ${\mathsf{LB_{ZZ}^M}}(P_\Xm, P_{\Ym|\Xm})$ in~\eqref{eq:discrete_zero_proof1} to be zero is to have $ P_e \left(\xv; \{ f_j \}_j, \{ \uv_j\}_j , M \right)$ in~\eqref{eq:discrete_zero_proof3} equal to zero almost surely for $t \geq 0$. This would indeed lead the integral in~\eqref{eq:discrete_zero_proof1} to be zero.

We note that, given  $t \geq 0$ and $\xv  \in \mathcal{X}$,  a sufficient condition for $   P_e \left(\xv;\Pc_\Uc(\xv),\Uc \right)$ to be zero is that there exists a $j \!\in\! [0:\mathsf{M}-1]$ for which $p_j\!=\! 1$, where $p_j\in\Pc_\Uc(\xv)$. Under this condition,  in fact, the optimal decision rule would always declare the correct hypothesis leading to a zero error probability. 

From the definition of $p_i\in\Pc_\Uc(\xv)$ in Theorem~\ref{thm:BZZ_Mary_vec_mmse}, we have
%for a discrete $\Xm$,  we have that
\begin{equation}
	p_i =\frac{  p_{\Xm-\uv_{i} }(\xv) }{   \sum_{{k=0}}^{\mathsf{M}-1}  p_{\Xm-\uv_{k} }(\xv) }, 
\end{equation}
and hence, there exists a $j \in [0:\mathsf{M}-1]$ for which $p_j  = 1$ if and only if $\bigcup_{\Ic\subseteq[0:\mathsf{M}-1]:|\Ic|\geq2} \bigcap_{k\in\Ic} \{\Xc - \uv_k\} = \varnothing$.

Thus, we have the following inclusion: for a given $\xv  \in \mathcal{X}$ 
\begin{equation}
\Ac \subseteq  \left\{ t \ge 0:   P_e \left(\xv;\Pc_\Uc(\xv),\Uc \right) = 0  \right\} ,
\end{equation} 
where 
\begin{align}\label{eq:discrete_zero_proof4}
	\Ac = \biggl\{t\geq0 : 
	& \bigcup_{\substack{\Ic\subseteq[0:\mathsf{M}-1] : \\ |\Ic| \geq 2}} \bigcap_{k\in\Ic} \{\Xc - \uv_k\} = \varnothing, \uv_k\in\Xc,  \nonumber \\ 
	& (\uv_k)_i = kt, ~k\in[0:\mathsf{M}-1] \biggr \}.
\end{align}
We now want to show that $\mathcal{A}$ above has a full measure, i.e., the complement of $\Ac$ on $[0,\infty)$ has Lebesgue measure zero. 
To this end, in Lemma~\ref{lem:discrete_zero_lemma}  in Appendix~\ref{app:discrete_zero_lemma}, we show that  $\mathcal{A}^c$ (i.e., the complement of $\mathcal{A})$ is countable and hence, $\mathcal{A}^c$ is of measure zero as desired.  
%\begin{lemma}\label{lem:discrete_zero_lemma}
%%For any discrete $\Xm$ with its sample space $\Xc$, the following set is countable:
%Assume that $\Xc$ is countable. Then, it holds that
%\begin{align}
%	\Bc = \biggl\{t\geq0 : 
%	& \bigcup_{\substack{\Ic\subseteq[0:M-1] \\ |\Ic| \geq 2}}  \bigcap_{k\in\Ic} \{\Xc - \uv_k\} \neq \varnothing, \uv_k\in\Xc, \nonumber \\
%	& \left<\ev_i, \uv_k\right> = kt, k\in[0:M-1] \biggr \}
%\end{align}
%is countable.
%\end{lemma}
%From Lemma~\ref{lem:discrete_zero_lemma}, it follows that $\mathcal{A}$ in~\eqref{eq:discrete_zero_proof4} has a full measure and hence, $ P_e \left(\xv; \{ f_j \}_j, \{ \uv_j\}_j , M \right)$ in~\eqref{eq:discrete_zero_proof3} is equal to zero almost surely for $t \geq 0$. 
This implies that $P_e \left(\xv;\Pc_\Uc(\xv),\Uc \right)$ in~\eqref{eq:discrete_zero_proof3} is equal to zero almost surely for $t > 0$ and hence,  the integral in~\eqref{eq:BZZ_Mary_vec_mmse_sub2} is equal to zero, leading to ${\mathsf{ZZ}}(P_\Xm, P_{\Ym|\Xm},\mathsf{M}) = 0$. This concludes the proof of Proposition~\ref{prop:discrete_zero_zzb}. 
%the integral in~\eqref{eq:discrete_zero_proof1} is equal to zero, which concludes the proof. 
\end{IEEEproof}

To show the remaining two results in the series, it is sufficient to focus on the case of a scalar input in the  high-noise regime,
i.e., to consider $\overline{\mathsf V}(P_X, \mathsf{M}) $ characterized in Theorem~\ref{thm:zzb_sig_inf_M}.
%Consider the quantity $\overline{\mathsf V}(P_X, \mathsf{M}) $ characterized in Theorem~\ref{thm:zzb_sig_inf_M}, and for scalar discrete inputs it is given by 
%\begin{align}\label{eq:Discrete_not_tight_proof1}
%	& \overline{\mathsf V}(P_X,\mathsf{M}) \nonumber \\
%	& =  \int_0^\infty \frac{t}{2} \Vc_t\left\{ \frac{ \mathsf{M} - \sum_{x\in\overline{\Xc}} \max_{k\in[0:\mathsf{M}-1]} p_X(x+kt) }{\mathsf{M}-1} \right\} {\rm d} t,
%\end{align}
%where $\overline{\Xc} = \cup_{k\in[0:M-1]} \{\Xc - kt\}$ and $p_X$ is the pmf of $X$. 
The next example,  the proof of which is in Appendix~\ref{app:ex_discrete_ber}, shows that, in general, $\overline{\mathsf V}(P_X,\mathsf{M}) \neq 0$.

\noindent {\bf{Example.}}
%\begin{example}\label{ex:discrete_ber}
Let $X\sim{\rm Ber}(p)$ be a Bernoulli random variable with parameter $0<p<1$. 
Then, 
\begin{equation}
\label{eq:Bern}
	\overline{\mathsf V}(P_X,2) =  \frac{1}{4}\min\{p,1-p\}.  
\end{equation}
%\end{example}

%As stated above, the valley-filling function in the {\color{cyan}ZZB} is necessary to obtain a proper bound for a discrete $X$. Next, we show that for a discrete $X$ the {\color{cyan}ZZB} with the valley-filling function is always loose (in high-noise regime).

The next result shows that even with the valley-filling function the ZZB is strictly sub-optimal. 
\begin{theorem}\label{thm:Discrete_not_tight}
Let $X$ be a discrete random variable  such that $\EE[X] = 0$ and $\inf\limits_{x\in\Xc}|x| >0$.
%i.e., there is no accumulation point at $0$.
Then, for any $\mathsf{M}$, it holds that
\begin{align}
	\overline{\mathsf V}(P_X, \mathsf{M}) < \var(X).
\end{align}
%Consequently, {\color{cyan}ZZB} for any discrete $X$ is not tight to the MMSE in high-noise regime.
\end{theorem}
\begin{IEEEproof}
An alternative form of $\var(X)$ is given by
\begin{align}\label{eq:Discrete_not_tight_proof2}
	\var(X)
	= \EE[X^2] 
	& = \int_0^\infty \frac{t}{2} \Pr\left(|X| \geq \frac{t}{2} \right) \ {\rm d} t.
\end{align}
Now, recall 
%from Theorem~\ref{thm:BZZ_Mary_vec_mmse} 
that  the ZZB was derived by  establishing the following lower bound: for $t >0$,
\begin{equation}
\Pr\left(|X| \!\geq \!\frac{t}{2} \right)  \!\ge \! \Vc_t\left\{  \frac{ \mathsf{M} - \sum_{x\in\overline{\Xc}} \max_{k\in[0:\mathsf{M}-1]} p_X(x+kt) }{\mathsf{M}-1} \right\},  \label{eq:Discrete_not_tight_proof3}
\end{equation} 
where $\overline{\Xc} = \bigcup_{j=0}^{\mathsf{M}-1} \left\{ \Xc - jt \right \}$.
We next show that~\eqref{eq:Discrete_not_tight_proof3} does not hold with equality and thus,  $ \var(X) > \overline{\mathsf V}(P_X, \mathsf{M})$.
% variance of $X$ is not equal to $\overline{\mathsf V}(P_X, \mathsf{M})$.  
First, let $x_0 = \inf_{x\in\Xc} |x|$ and note that the left-hand side of~\eqref{eq:Discrete_not_tight_proof3} is
\begin{equation}\label{eq:Discrete_not_tight_proof33}
	\Pr\left(|X| \geq \frac{t}{2} \right) 
	= 1, \text{ if } t \leq x_0.
\end{equation}
Second,  assume $t>0$ and consider the sum inside the right-hand side of~\eqref{eq:Discrete_not_tight_proof3}. 
Since $p_X(x+kt) = p_{X-kt}(x)$, we have
\begin{equation}\label{eq:Discrete_not_tight_proof5}
	\sum_{x\in\overline{\Xc}} \max_{k\in[0:\mathsf{M}-1]} p_X(x+kt) 
	\overset{\rm (a)}{>} \sum_{x\in{\Xc}} \max_{k\in[0:\mathsf{M}-1]} p_{X-kt}(x)
	\overset{\rm (b)}{\geq} 1,
\end{equation}
where $\rm (a)$ follows by the fact that $\overline{\Xc}\setminus \Xc \neq \varnothing$ if $t>0$, and $\rm (b)$ follows by dropping the $\max$ and choosing $k=0$.
Note that~\eqref{eq:Discrete_not_tight_proof5} is true for any $t > 0$ and hence,  from the definition of valley-filling function in~\eqref{eq:VF}, it follows that the right-hand side of~\eqref{eq:Discrete_not_tight_proof3} is strictly smaller than one for all $t>0$.  This, together with~\eqref{eq:Discrete_not_tight_proof33}, shows that
there exists a range of $t$, namely $ 0< t \leq x_0$, such that the condition in~\eqref{eq:Discrete_not_tight_proof3} does not hold with equality.
This concludes the proof of Theorem~\ref{thm:Discrete_not_tight}.
\end{IEEEproof}

\begin{remark}
In Theorem~\ref{thm:Discrete_not_tight}, the assumption that the support of $X$ has no accumulation point at zero was done to make the proof easier and most likely can be removed. 
Moreover, the result in Theorem~\ref{thm:Discrete_not_tight} shows that, when working with discrete inputs, the ZZB might not be the best to use, especially in the practically relevant high-noise regime.  
\end{remark}

\section{Conclusion}
This paper has provided an expression for a general Ziv-Zakai bound, which requires no regularity assumptions. In particular, we have removed the continuity assumption, and the bound now holds for all input distributions. The paper has also shown that care must be taken when dealing with discrete distributions and that the valley-filling function is a necessary tool for this case.  The paper has also characterized the high-noise and low-noise asymptotics.  The first key observation here is that in the low-noise regime, the bound is tight in its simplest form (i.e., without the valley-filling function, and the number of hypotheses is taken to be two) under additive Gaussian noise. The second key observation is that in the high-noise regime, the situation is more complex, and the bound might not be tight.  For continuous distributions, the paper has provided the necessary and sufficient conditions for tightness. Finally, for discrete inputs, the paper has shown that the bound, in general, is sub-optimal in the high-noise regime. The last observation suggests that, for discrete inputs, either a new bound is needed or a new modification of the Ziv-Zakai bound is needed to make it tighter.

\newpage

\bibliographystyle{IEEEtran}
\bibliography{ISIT2023_arxiv}

\newpage

 \appendices
%%
%% in combination with further \section-commands can be used.
%%%%%%

\section{Proof of Theorem~\ref{thm:BZZ_Mary_vec_mmse} }\label{app:gen_zzb}
The proof of Theorem~\ref{thm:BZZ_Mary_vec_mmse} is a direct consequence of the following Theorem~\ref{thm:BZZ_Mary_vec}.
In particular, since Theorem~\ref{thm:BZZ_Mary_vec} holds for any estimator $g(\Ym) \in \mathcal{X}$ and any {$\av\in \mathcal{X}$,} we can obtain a lower bound on ${\rm mmse}(\Xm|\Ym)$ by taking $g(\Ym) = \EE[\Xm|\Ym]$ and $\av = \ev_i$ in Theorem~\ref{thm:BZZ_Mary_vec}  and by summing it over $i\in[1:\mathsf{d}(\Xc)]$. This would lead us to the MMSE lower bound in Theorem~\ref{thm:BZZ_Mary_vec_mmse}.
\begin{theorem}
%~\cite[eq.~(3.85), ep.~(3.86)]{bell1995performance}
\label{thm:BZZ_Mary_vec}
%{\color{red}Let $\mathcal{X}$ be a finite dimensional real-valued vector space.}  
Let $\mathcal{X}$ be a vector space.
Let $\epsilonv = g(\Ym) - \Xm$ be the estimation error when the estimator $g(\Ym)\in \mathcal{X}$ is used to estimate $\Xm\in \mathcal{X}$.
Then, for every $ \av \in \mathcal{X}$ and any integer $\mathsf{M}\geq2$, we have that
\begin{equation}\label{eq:thmBBZ_Mary_vec}
	\EE\left[  \left< \av, \epsilonv \right>^2 \right]
	\geq \int_0^\infty \frac{t}{2} \Vc_t \left\{ \frac{{\mathsf{h}(t,\av,\mathsf{M})}}{\mathsf{M}-1} \right\} \ {\rm d} t,
\end{equation}
where, for $ t >0$ and $ i \in [0:\mathsf{M}-1]$, it holds that 
 \begin{equation}
	\mathsf{h}(t,\av,\mathsf{M}) 
	\! = \!\! \max\limits_{\substack{  \mathcal{U}  \subset \mathcal{X}: \\ \mathcal{U}= \{ \uv_k \}_{k=0}^{\mathsf{M}-1}  \\  \left< \av, \uv_k \right> = kt, \\ k \in [0:\mathsf{M}-1] }}  \, \int    P_e \left ({\xv}; \mathcal{P}_\mathcal{U}(\xv), \mathcal{U}  \right )  \,    {\rm{d}} \mu_{\mathcal{U}} (\xv)    .
\end{equation}
To define $\mathsf{h}(t,\av,\mathsf{M})$, we let:
\begin{itemize}[leftmargin=*]
\item for a set $\mathcal{U}= \{ \uv_k \}_{k=0}^{\mathsf{M}-1} \subset \mathcal{X} $, the measure $\mu_\Uc$ is given by
\begin{equation}
\label{eq:def_mu_U}
	\mu_{\mathcal{U}} =\sum_{{k=0}}^{\mathsf{M}-1}  P_{\Xm-\uv_{k} }; 
\end{equation} 
\item   for $\xv \in \mathcal{X}$ and $\mathcal{U} \subset \mathcal{X}$, we let 
 \begin{equation}
 \label{eq:def_P_U}
 \mathcal{P}_\mathcal{U}(\xv) = \left \{ p_i: p_i =  \frac{  {\rm{d}} P_{\Xm-\uv_{i} } }{   {\rm{d} \mu_{\mathcal{U}}}   } (\xv),   \uv_i \in \mathcal{U}= \{ \uv_k \}_{k=0}^{\mathsf{M}-1}
 \right \};
 \end{equation} 
note that by construction, the Radon-Nikodym derivative is always well defined.
% since  $P_{\Xm-\uv_{i} }(\cdot)$ is absolutely continuous with respect to  $\mu_{\mathcal{U}}$.
% \item for $ t >0$ and $ i \in [0:M-1]$, we let 
\end{itemize} 
\end{theorem}

\begin{IEEEproof}
We start by defining a few terms. First, note that
\begin{align}\label{eq:BZZ_Mary_vec_proof1}
	\EE[  \left< \av, \epsilonv \right>^2 ]
	 = \int_0^\infty \frac{t}{2} \Pr\left( |\left< \av, \epsilonv \right>| \geq \frac{t}{2} \right) \ {\rm d} t.
\end{align}
Second, fix some $\mathsf{M} \ge 2$,  choose a  set of vectors $\Uc = \{ \uv_i \}_{i=0}^{\mathsf{M}-1} \subset \mathcal{X}$, and define $\mathcal{P}_\mathcal{U}(\xv)$ as in~\eqref{eq:def_P_U}.
%the following set,
% \begin{equation}
 %\mathcal{P}_\mathcal{U}(\xv) = \left \{ p_i: p_i =  \frac{  {\rm{d}} P_{\Xm-\uv_{i} } }{   {\rm{d} \mu_{\mathcal{U}}}   } (\xv),   \uv_i \in \mathcal{U}= \{ \uv_k \}_{k=0}^{M-1}
 %\right \} , 
% \end{equation} 
 %where 
 %\begin{equation}\label{eq:def_mu_U}
%\mu_{\mathcal{U}} =\sum_{{k=0}}^{M-1}  P_{\Xm-\uv_{k} }.
%\end{equation} 
%\begin{align}
%f_i \left (\xv;  \{ \uv_i \}_{i=0}^{M-1} \right)= \frac{  {\rm{d}} P_{\Xm-\uv_{i} }(\xv) }{ \sum_{k=0}^{M-1} {\rm{d}} P_{\Xm-\uv_{k} }(\xv) },  \, i \in \{0, \ldots, M-1 \}. \label{eq:def_f_i}
%\end{align}
%We note that  the Radon-Nikodym derivative $p_i$ is well-defined since  $P_{\Xm-\uv_{i} }(\cdot)$ is absolutely continuous with respect to  $\sum_{{k=0}}^{M-1}  P_{\Xm-\uv_{k}}(\cdot)$.
Next, let $\left<\av, \uv_i\right> = it$, and note that 
\begin{align}\label{eq:gen_ZZB_proof1}
& \Pr\left( |\left< \av, \epsilonv \right>| \geq \frac{t}{2} \right) \nonumber \\
%
%& =  \Pr\left( \left< \av, \epsilonv \right> \geq \frac{t}{2} \right)  + \Pr\!\left( \left< \av, \epsilonv \right> \leq - \frac{t}{2} \right) \nonumber \\
%
%& = \! \left( \! \Pr \! \left( \left< \av, \epsilonv \right> \geq \frac{t}{2} \right)  \! + \Pr\!\left(\!\left< \av, \epsilonv \right> \leq - \frac{t}{2} \right) \! \right)\\
%
 = &  \int {\rm{d}} P_{\Xm}(\xv) \Pr \left( \left< \av, g(\Ym) - \Xm \right>  \!\geq \!  \frac{t}{2} \; \middle | \; \Xm=\xv  \right)   \nonumber \\
& \quad +   \int {\rm{d}} P_{\Xm}(\xv)   \Pr \left( \left< \av, g(\Ym) - \Xm \right> \! \leq \! - \frac{t}{2} \; \middle | \; \Xm=\xv \right) \nonumber \\
=& \frac{1}{\mathsf{M}-1} \sum_{i=0}^{\mathsf{M}-1} \left \{  \int {\rm{d}} P_{\Xm}(\xv) \Pr \left( \left< \av, g(\Ym) -\Xm \right>  \!\geq \! \frac{t}{2} \; \middle | \; \Xm=\xv  \right)  \right. \nonumber \\
& \quad \left. +\!   \int \! {\rm{d}} P_{\Xm}(\xv)   \Pr \left( \left< \av, g(\Ym) \!-\! \Xm \right> \! \leq \! - \frac{t}{2} \; \middle | \; \Xm\!=\!\xv \right)  \right\} ,
\end{align}
where the first equality follows by applying the law of total probability and substituting $\epsilonv = g(\Ym) - \Xm$.
The summation in~\eqref{eq:gen_ZZB_proof1} is then given by~\eqref{eq:Getting_to_the_error_prob_step}, at the top of the next page,
\begin{figure*}
\begin{align}\label{eq:Getting_to_the_error_prob_step}
& \sum_{i=0}^{\mathsf{M}-1} \left \{  \int {\rm{d}} P_{\Xm}(\xv) \Pr \left( \left< \av, g(\Ym) -\Xm \right>  \!\geq \! \frac{t}{2} \; \middle | \; \Xm=\xv  \right)  +   \int {\rm{d}} P_{\Xm}(\xv)   \Pr \left( \left< \av, g(\Ym) -\Xm\right> \! \leq \! -\frac{t}{2} \; \middle | \; \Xm=\xv \right)  \right\} \notag \\
%
%& \stackrel{{\rm{(c)}}}{=}  \sum_{i=1}^{M-1} \left \{  \int {\rm{d}} P_{\Xm }(\xv+\uv_{i-1}) \Pr \left( \left< \av, g(\Ym) -\Xm \right>  \!\geq \!  \frac{t}{2} \; \middle | \; \Xm=\xv+\uv_{i-1}  \right)    \right. \notag\\
%&\qquad \qquad  +\left.    \int {\rm{d}} P_{\Xm}(\xv+\uv_{i})   \Pr \left( \left< \av, g(\Ym) -\Xm \right> \! \leq \! - \frac{t}{2} \; \middle | \; \Xm=\xv+\uv_{i} \right)  \right\} \notag\\
%
& \stackrel{{\rm{(a)}}}{=}  \sum_{i=0}^{\mathsf{M}-1} \left \{  \int {\rm{d}} P_{\Xm-\uv_{i-1} }(\xv) \Pr \left( \left< \av, g(\Ym) \!-\! \Xm \right>  \!\geq \! \frac{t}{2} \; \middle | \; \Xm=\xv+\uv_{i-1}  \right)    \!+\!   \int {\rm{d}} P_{\Xm-\uv_{i}}(\xv)   \Pr \left( \left< \av, g(\Ym) \!-\! \Xm \right> \! \leq \! - \frac{t}{2} \; \middle | \; \Xm=\xv+\uv_{i} \right)  \right\} \notag \\
&\stackrel{{\rm{(b)}}}{=}   \sum_{i=0}^{\mathsf{M}-1} \left \{  \int {\rm{d}} P_{\Xm-\uv_{i-1} }(\xv) \Pr \left( \left< \av, g(\Ym) \!-\! \xv \right>  \!\geq \! {\left( i-\frac{1}{2} \right)}t\; \middle | \; \Xm=\xv+\uv_{i-1}  \right)    \right. \notag\\
& \qquad \qquad +\left .    \int {\rm{d}} P_{\Xm-\uv_{i}}(\xv)   \Pr \left( \left< \av, g(\Ym) \!-\! \xv \right> \! \leq \!  \left( i-\frac{1}{2} \right)t  \; \middle | \; \Xm=\xv+\uv_{i} \right)  \right\} \notag \\
&\stackrel{{\rm{(c)}}}{=}   \sum_{i=0}^{\mathsf{M}-1} \left \{  \int  p_{i-1} \Pr \left( \left< \av, g(\Ym) \!-\! \xv \right>  \!\geq \!  \left( i-\frac{1}{2} \right)t\; \middle | \; \Xm=\xv+\uv_{i-1}  \right)  \ {\rm d}\mu_\Uc(\xv)  \right. \notag\\
& \qquad \qquad +\left .    \int  p_i   \Pr \left( \left< \av, g(\Ym) \!-\! \xv \right> \! \leq \! \left( i-\frac{1}{2} \right)t  \; \middle | \; \Xm=\xv+\uv_{i} \right) \ {\rm d}\mu_\Uc(\xv) \right\} \notag \\
&\stackrel{{\rm{(d)}}}{=} \int  P_e^\phi \left(\xv,t ; \Pc_\Uc(\xv), \Uc \right) \ {\rm d}\mu_\Uc(\xv),
\end{align}
\hrule
\end{figure*}where the labeled equalities follow from:
%$\rm{(a)}$ applying the law of total expectation;
%$\rm{(b)}$ substituting $\epsilonv = g(\Ym) - \Xm$ and using the linearity of the inner product;
$\rm{(a)}$ applying the changes of variables $\xv^\prime = \xv +\uv_{i-1}$ in the first integral and $\xv^\prime = \xv +\uv_{i}$ in the second integral, and the fact that ${\rm{d}} P_{\Xm}(\xv+\uv_{i-1}) = {\rm{d}} P_{\Xm-\uv_{i-1} }(\xv)$;
 $\rm{(b)}$ using the assumption that $\left<\av, \uv_i\right> = it$;
 $\rm{(c)}$  using $p_i\in\Pc_\Uc(\xv)$ in~\eqref{eq:def_P_U} and $\mu_\Uc$ in~\eqref{eq:def_mu_U};
 and $\rm{(d)}$ defining $P_e^\phi \left(\xv ; \Pc_\Uc(\xv), \Uc \right)$ as the error probability for an $\mathsf{M}$-ary hypothesis testing problem as in Definition~\ref{def:MHT} associated with the (possibly sub-optimal) decision rule,
\begin{align}
	\phi = \Hc_i, \text{ where } i = \argmin_{ j \in [0:\mathsf{M}-1] } | \left< \av , g(\Ym) -\xv \right> - jt |.
\end{align}
%\begin{align}
%&P_e^\phi \left(\xv,t ; \Pc_\Uc(\xv), \Uc \right) \notag\\
%&= p_0  \Pr \left( \left< \av, g(\Ym) \! - \! \xv \right>  \!\geq \!  \frac{t}{2} \; \middle | \; \Xm=\xv+\uv_{0}  \right)  \notag\\
%&\quad+  \sum_{i=1}^{M-2} p_i  \left(\Pr \left( \left< \av, g(\Ym) \! - \! \xv \right> \! \geq \! \left( i+\frac{1}{2} \right)t  \; \middle | \; \Xm=\xv+\uv_{i} \right)  + \Pr \left( \left< \av, g(\Ym) \! - \! \xv \right> \! \leq \!  \left( i-\frac{1}{2} \right)t  \; \middle | \; \Xm=\xv+\uv_{i} \right)  \right)  \notag\\
%& \quad   + p_{M-1}     \Pr \left( \left< \av, g(\Ym) \!-\! \xv \right> \! \leq \!  \left( M-\frac{3}{2} \right)t  \; \middle | \; \Xm=\xv+\uv_{M-1} \right)  . 
%\end{align}
%We note that $P_e^\phi \left(\xv,t ; \Pc_\Uc(\xv), \Uc \right) $  is  the probability of error associated with the following (possibly sub-optimal) decision rule,
%\begin{align}
%	\phi = 
%	\begin{cases}
%		H_0 & \text{ if } \left< \av , g(\Ym) -\xv \right> <  \frac{t}{2} \\
%		H_i & \text{ if } \left(i - \frac{1}{2}\right)t < \left< \av , g(\Ym) -\xv \right>  < \left( i + \frac{1}{2}\right)t \\
%		H_{M-1} & \text{ if } \left( M-\frac{3}{2}\right)t < \left< \av , g(\Ym) -\xv \right> 
%	\end{cases}.
%\end{align}
Combining~\eqref{eq:gen_ZZB_proof1} and~\eqref{eq:Getting_to_the_error_prob_step},
% and replacing $\phi$ with the optimal decision rule, 
 we obtain
\begin{align}
	\Pr\left( |\left< \av, \epsilonv \right>| \geq \frac{t}{2} \right)
	&=\frac{\int   P_e^\phi \left(\xv ; \Pc_\Uc(\xv), \Uc \right) \ {\rm d}\mu_\Uc(\xv)}{\mathsf{M}-1}  \notag\\
	& \ge \frac{ \int   P_e \left(\xv; \Pc_\Uc(\xv), \Uc \right)  \ {\rm d}\mu_\Uc(\xv) }{\mathsf{M}-1},
\label{eq:Before_valley}
\end{align}
where we have used the fact that $P_e \left(\xv ; \Pc_\Uc(\xv), \Uc \right) $ is the error probability associated with the optimal decision rule (i.e., the MAP decision rule). The lower bound in~\eqref{eq:Before_valley} can be further tightened by optimizing $\Uc = \{\uv_i\}_{i=0}^{\mathsf{M}-1}$ such that $\left<\av,\uv_i\right> = it,~\forall i\in[0:\mathsf{M}-1]$.
Then,
\begin{align}\label{eq:Before_valley2}
	\Pr\left( |\left< \av, \epsilonv \right>| \geq \frac{t}{2} \right)
	& \geq \max\limits_{\substack{  \mathcal{U}  \subset \mathcal{X}: \\ \mathcal{U}= \{ \uv_k \}_{k=0}^{\mathsf{M}-1}  \\  \left< \av, \uv_k \right> = kt, \\ k \in [0:\mathsf{M}-1] }}  \, \frac{ \int    P_e \left ({\xv}; \mathcal{P}_\mathcal{U}(\xv), \mathcal{U}  \right )  \, {\rm{d}} \mu_{\mathcal{U}} (\xv) }{\mathsf{M}-1}  \nonumber \\
	& = \frac{ \mathsf{h}(t,\av,\mathsf{M}) }{\mathsf{M}-1}.
\end{align}
Applying the valley-filling function to the right-hand side of~\eqref{eq:Before_valley2} is valid since $\Vc_t\{\Pr\left( {\left | \left< \av,  \epsilonv \right> \right |} \geq \frac{t}{2} \right)\} = \Pr\left( {\left | \left< \av,  \epsilonv \right> \right |} \geq \frac{t}{2} \right)$ due to the monotonicity of $\Pr\left( {\left | \left< \av,  \epsilonv \right> \right |} \geq \frac{t}{2}\right)$ with respect to $t$. Thus, we obtain 
\begin{align}
\label{eq:BZZ_Mary_scalar_proof7}
	& \Pr\left( {\left | \left< \av,  \epsilonv \right> \right |} \geq \frac{t}{2} \right)   \geq \Vc_t \left\{ \frac{\mathsf{h}(t,\av,\mathsf{M})}{\mathsf{M}-1}  \right\}. 
\end{align}
Substituting~\eqref{eq:BZZ_Mary_scalar_proof7} into~\eqref{eq:BZZ_Mary_vec_proof1} finishes the proof of Theorem~\ref{thm:BZZ_Mary_vec}.
\end{IEEEproof}

\section{Proof of Theorem~\ref{thm:zzb_sig_inf_M}}\label{app:high_noise}

At first, we observe that if $f(t)\leq g(t)$ for all $t\in\RR$, then 
\begin{align}\label{eq:valley_upper}
	\Vc_t\{f(t)\} \leq \Vc_t\{g(t)\},~\forall t\in\RR.
\end{align}
Second, the valley-filling function is lower semicontinuous, i.e., we have that
\begin{align}\label{eq:valley_liminf}
	\liminf_{n\to\infty} \Vc_t\{ f_n(t)\} \geq \Vc_t\{\liminf_{n\to\infty} f_n(t)\}.
\end{align}
To highlight the dependency of $\mathsf{h}(t,i,\mathsf{M})$ in~\eqref{eq:BZZ_Mary_vec_mmse_h} on $\eta$, we use $\mathsf{h}_\eta(t,i,\mathsf{M})$.
We also note that $\mathsf{h}_\eta(t,i,\mathsf{M})$ is non-decreasing with respect to the noise level $\eta$ since $ P_e \left (\eta, {\xv}; \mathcal{P}_\mathcal{U}(\xv), \mathcal{U}  \right )  $ is non-decreasing in $\eta$ as assumed in {\bf A2}.
%As introduced the noise level $\eta$, we highlight the dependency of $\mathsf{h}(t,i,M)$ in~\eqref{eq:BZZ_Mary_vec_mmse_h} on $\eta$ by writing it as $\mathsf{h}_\eta(t,i,M)$.
%Note that the function $\mathsf{h}(t,i,M)$ in~\eqref{eq:BZZ_Mary_vec_mmse_h} is non-decreasing with respect to the noise level $\eta$ since $ P_e \left (\eta, {\xv}; \mathcal{P}_\mathcal{U}(\xv), \mathcal{U}  \right )  $ is non-decreasing in $\eta$ as assumed in {\bf A2}.
We then have that the Ziv-Zakai lower bound in~\eqref{eq:BZZ_Mary_vec_mmse_sub1} can be written as
\begin{align}\label{eq:BZZ_Mary_vec_mmse_sub1_proof}
	\overline{\mathsf{ZZ}}(P_\Xm, P_{\Ym|\Xm},\mathsf{M})
	& = \sum_{i=1}^{\mathsf{d}(\Xc)} \!\! \int_0^\infty \frac{t}{2} \Vc_t\left\{ \frac{\mathsf{h}_\eta(t,i,\mathsf{M})}{\mathsf{M}-1} \right\} \ {\rm d} t \nonumber \\
	& \leq \!\sum_{i=1}^{\mathsf{d}(\Xc)} \!\! \!\int_0^\infty \frac{t}{2} \Vc_t\left\{ \frac{\mathsf{h}_\infty(t,i,\mathsf{M})}{\mathsf{M}-1} \right\} \ {\rm d} t.
\end{align}
For the sake of space, we abbreviate the constraint $ \{ \mathcal{U}= \{ \uv_k \}_{k=0}^{\mathsf{M}-1} , (\uv_k )_i = kt,~\forall k \}$ as $\Cc$.
Due to the assumption {\bf A3}, we can write $\mathsf{h}_\infty(t,i,\mathsf{M})$ as
\begin{align}\label{eq:BZZ_Mary_vec_mmse_h_infty}
	\mathsf{h}_\infty  (t,i,\mathsf{M}) 
	& = \max\limits_{\substack{  \mathcal{U}  \subset \mathcal{X}: \Cc }}  \int    P_e \left (\infty, {\xv}; \mathcal{P}_\mathcal{U}(\xv), \mathcal{U}  \right )  \ {\rm d}\mu_\Uc(\xv) \nonumber\\
	& = \max\limits_{\substack{  \mathcal{U}  \subset \mathcal{X}: \Cc }} \left\{  \mathsf{M} - \int  \max_{j\in[0:\mathsf{M}-1]}  {\rm{d}} P_{\Xm-\uv_{j} }(\xv) \right\} \nonumber \\
	& =   \mathsf{M}  - \min\limits_{\substack{  \mathcal{U}  \subset \mathcal{X}: \Cc }}  \int  \! \! \max_{j\in[0:\mathsf{M}-1]}  {\rm{d}} P_{\Xm-\uv_{j} }(\xv)  ,
\end{align}
%where the second equality follows by the assumption {\bf A3}.
which,  substituted inside~\eqref{eq:BZZ_Mary_vec_mmse_sub1_proof}, leads to
\begin{align}
\label{eq:zzb_sig_inf_upper}
	& \overline{\mathsf{ZZ}}(P_\Xm, P_{\Ym|\Xm},\mathsf{M}) \nonumber \\
	& \leq \sum_{i=1}^{\mathsf{d}(\Xc)} \int_0^\infty \! \frac{t}{2} \Vc_t  \left\{\! \frac{\mathsf{M}  -\mathsf{H}(P_\Xm, t,i,\mathsf{M})}{\mathsf{M}-1} \!\right\} {\rm d} t.
\end{align}
%
%\begin{align}\label{eq:zzb_sig_inf_upper}
%	\overline{\mathsf{LB_{ZZ}}}(\Xm,\eta)
%	& = \sum_{i=1}^n \int_0^\infty \frac{t}{2} G\left( \frac{1}{M-1} \max_{\substack{ \uv_0,...,\uv_{M-1} \\ \ev_i^{\mathsf{T}} \uv_k = kt,~ \forall k}} \int \left( \sum_{j=0}^{M-1} f_\Xm(\xv+\uv_j) \right) P_e(\eta,\{\xv+\uv_i\}_0^{M-1}, \{p_i\}_0^{M-1}) {\rm d} \xv \right) {\rm d} t \nonumber \\
%	& \overset{\rm (a)}{\leq} \sum_{i=1}^n \int_0^\infty \frac{t}{2} G\left( \frac{1}{M-1} \max_{\substack{ \uv_0,...,\uv_{M-1} \\ \ev_i^{\mathsf{T}} \uv_k = kt,~ \forall k}} \int \left( \sum_{j=0}^{M-1} f_\Xm(\xv+\uv_j) \right) P_e(\infty,\{\xv+\uv_i\}_0^{M-1}, \{p_i\}_0^{M-1}) {\rm d} \xv \right) {\rm d} t \nonumber \\
%	& \overset{\rm (b)}{=} \sum_{i=1}^n \int_0^\infty \frac{t}{2} G\left( \frac{1}{M-1} \max_{\substack{ \uv_0,...,\uv_{M-1} \\ \ev_i^{\mathsf{T}} \uv_k = kt,~ \forall k}} \int \left( \sum_{j=0}^{M-1} f_\Xm(\xv+\uv_j) \right) (1 -  \max_{i\in[0:M-1]} p_i ) {\rm d} \xv \right) {\rm d} t \nonumber \\
%	& = \sum_{i=1}^n \int_0^\infty \frac{t}{2} G\left( \frac{1}{M-1} \max_{\substack{ \uv_0,...,\uv_{M-1} \\ \ev_i^{\mathsf{T}} \uv_k = kt,~ \forall k}} \left( M  -  \int  \max_{i\in[0:M-1]} f_\Xm(\xv+\uv_i) {\rm d} \xv  \right)  \right) {\rm d} t ,
%\end{align}
%where the labeled (in)equalities follow from:
%$\rm (a)$ the assumption A2a;
%$\rm (b)$ the assumption A2b.
Next, we have that
\begin{align}\label{eq:BZZ_Mary_vec_mmse_sub1_lower}
	& \liminf_{\eta\to\infty}   \overline{\mathsf{ZZ}}(P_\Xm, P_{\Ym|\Xm},\mathsf{M}) \nonumber
	\\& \overset{\rm (a)}{\geq} \!\sum_{i=1}^{\mathsf{d}(\Xc)}  \int_0^\infty \!\! \liminf_{\eta\to\infty}  \frac{t}{2} \Vc_t \left\{ \! \frac{\mathsf{h}_\eta(t,i,\mathsf{M})}{\mathsf{M}-1} \! \right\} \ {\rm d} t \nonumber \\
	& \overset{\rm (b)}{\geq} \sum_{i=1}^{\mathsf{d}(\Xc)}  \int_0^\infty \frac{t}{2} \Vc_t \left\{ \! \frac{ \liminf_{\eta\to\infty} \mathsf{h}_\eta(t,i,\mathsf{M})}{\mathsf{M}-1} \! \right\} \ {\rm d} t,
\end{align}
where $\rm (a)$ is due to Fatou's lemma, and $\rm (b)$ follows from~\eqref{eq:valley_liminf}.
Moreover,  we have that
\begin{align}\label{eq:BZZ_Mary_vec_mmse_h_liminf}
	\!\! \liminf_{\eta\to\infty} \mathsf{h}_\eta(t,i,\mathsf{M})   
	& \!=\!  \liminf_{\eta\to\infty} \!\! \max\limits_{\substack{  \mathcal{U}  \subset \mathcal{X}: \Cc }} \!  \int  \!\!  P_e \left (\eta, {\xv}; \mathcal{P}_\mathcal{U}(\xv), \mathcal{U}  \right ) \ {\rm d}\mu_\Uc(\xv) \nonumber \\
	& \! \overset{\rm (a)}{\geq} \! \max\limits_{\substack{  \mathcal{U}  \subset \mathcal{X}: \Cc }} \!\liminf_{\eta\to\infty} \! \! \int  \!\! P_e \left (\eta, {\xv}; \mathcal{P}_\mathcal{U}(\xv), \mathcal{U}  \right ) \ {\rm d}\mu_\Uc(\xv)  \nonumber \\
	& \! \overset{\rm (b)}{\geq} \max\limits_{\substack{  \mathcal{U}  \subset \mathcal{X}: \Cc }}  \int   P_e \left (\infty, {\xv}; \mathcal{P}_\mathcal{U}(\xv), \mathcal{U}  \right ) \  {\rm d}\mu_\Uc(\xv) \nonumber \\
%	& \overset{\rm (a)}{\geq} \max_{\substack{ \uv_k, \forall k \\ \ev_i^{\mathsf{T}} \uv_k = kt}}  \liminf_{\eta\to\infty}  \int   P_e(\eta,\{\xv + \uv_k\}_0^{M-1})  \sum_{j=0}^{M-1} {\rm d} P_\Xm(\xv+\uv_j)  \nonumber \\
%	& \overset{\rm (b)}{\geq} \max_{\substack{ \uv_k, \forall k \\ \ev_i^{\mathsf{T}} \uv_k = kt}}   \int  P_e(\infty,\{\xv + \uv_k\}_0^{M-1})  \sum_{j=0}^{M-1} {\rm d} P_\Xm(\xv+\uv_j) \nonumber \\
	&  \overset{\rm (c)}{=} \mathsf{M} \!-\!\! \min\limits_{\substack{  \mathcal{U}  \subset \mathcal{X}:\Cc }} \int  \!\!\!\!\max_{j\in[0:\mathsf{M}-1]}\!\!  {\rm{d}} P_{\Xm-\uv_{j} }(\xv),
\end{align}
where 
$\rm (a)$ follows from the max-min inequality, $\rm (b)$ is due to Fatou's lemma, and $\rm (c)$ is from the assumption {\bf A3}.
%, which is similar to the steps in~\eqref{eq:BZZ_Mary_vec_mmse_h_infty}.
Substituting~\eqref{eq:BZZ_Mary_vec_mmse_h_liminf} into~\eqref{eq:BZZ_Mary_vec_mmse_sub1_lower}, 
%we arrive at a lower bound when $\eta\to\infty$, that is, for $\eta\to\infty$ we have that
we obtain that, for $\eta\to\infty$
\begin{align*}
%\label{eq:zzb_sig_inf_lower}
%	\liminf_{\eta\to\infty}  \overline{\mathsf{LB_{ZZ}^M}}(P_\Xm, P_{\Ym|\Xm})
%	& \geq  
	 \overline{\mathsf{ZZ}}(P_\Xm, P_{\Ym|\Xm},\mathsf{M})
	 \geq \!\sum_{i=1}^{\mathsf{d}(\Xc)}  \!\int_0^\infty \frac{t}{2} \Vc_t\left\{ \frac{ \mathsf{M} \! -\! \mathsf{H}(P_\Xm,t,i,\mathsf{M})  }{\mathsf{M}-1} \right\} {\rm d} t,
\end{align*}
which agrees with \eqref{eq:zzb_sig_inf_upper} and completes the proof of Theorem~\ref{thm:zzb_sig_inf_M}.
%We have that \eqref{eq:zzb_sig_inf_upper} coincides with~\eqref{eq:zzb_sig_inf_lower}, and this concludes the proof of Theorem~\ref{thm:zzb_sig_inf_M}.

\section{Proof of Theorem~\ref{thm:low_noise}}\label{app:low_noise_proof}
%Consider $\mathsf{M}=2$ and note that proving $\mathsf{M}=2$ case is sufficient. 
Using the change of variable $t = \sqrt{\eta} \tau$ in~\eqref{eq:BZZ_Mary_vec_mmse_sub2},  we obtain
\begin{align}\label{eq:low_noise_proof0}
	\frac{ {\mathsf{ZZ}}(P_\Xm, P_{\Ym|\Xm},2)}{\eta} 
%	& = \sum_{i=1}^n  \int_0^\infty \frac{t}{2\sigma^2} h(t,\ev_i,2)  \, {\rm d} t \nonumber \\
	& = \sum_{i=1}^{\mathsf{d}(\Xc)}  \int_0^\infty \frac{\tau}{2} \mathsf{h}(\sqrt{\eta}\tau,i,2) \, {\rm d} \tau.
%	& = \sum_{i=1}^n  \int_0^\infty \frac{\tau}{2}   \max\limits_{\substack{  \mathcal{U}  \subset \mathcal{X}: \\ \mathcal{U}= \{ \vv_k \}_{k=0}^{1}  \\  \left< \ev_i, \vv_k \right> = k\tau, \\ k \in [0:1] }}  \, \int  \left(  \int_{\yv^\prime  \in \overline{\Rc}_0}   f_\Zm(\yv^\prime - \vv_1 ) f_\Xm(\xv+\sigma \vv_1)  {\rm  d} \yv^\prime  \right. \nonumber \\
%	&~~~~~~~~~~~~~~~  \left. +   \int_{\yv^\prime  \in \overline{\Rc}_1}  f_\Zm(\yv^\prime - \vv_0 ) f_\Xm(\xv+\sigma \vv_0)  {\rm  d} \yv^\prime  \,   \right) {\rm d}\xv   \, {\rm d} \tau.
\end{align}
Moreover, since we can write 
\begin{align}
	\mu_\Uc 
%	& = \sum_{k=0}^{M-1} P_{\Xm - \uv_k} \nonumber\\
	& = \alpha\sum_{k=0}^{1} P_{\Xm_C - \uv_k} + (1-\alpha)\sum_{k=0}^{1} P_{\Xm_D - \uv_k}  \nonumber\\
	& = \alpha \mu_{\Uc_C} + (1-\alpha)\mu_{\Uc_D},
\end{align}
we can lower bound $\mathsf{h}(\sqrt{\eta}\tau,i,2)$ in~\eqref{eq:low_noise_proof0} as
 \begin{align}\label{eq:low_noise_proof_cont}
	\mathsf{h}(\sqrt{\eta}\tau,i,2)
%	& = \!\! \max\limits_{\substack{  \mathcal{U}  \subset \mathcal{X}: \\ \mathcal{U}= \{ \uv_k \}_{k=0}^{1}  \\  \left< \ev_i, \uv_k \right> = \sigma k\tau, \\ k \in [0:1] }}  \, \int    P_e \left ({\xv}; \mathcal{P}_\mathcal{U}(\xv), \mathcal{U}  \right )  \,    {\rm{d}} \mu_{\mathcal{U}} (\xv)    \nonumber \\
	& \overset{\rm (a)}{\geq} \!\!  \max\limits_{\substack{  \mathcal{U}  \subset \mathcal{X}: \\ \mathcal{U}= \{ \uv_k \}_{k=0}^{1}  \\  ( \uv_k )_i = \sqrt{\eta} k\tau, ~\forall k}}  \!\!\!\! \alpha \int    P_e \left (\eta, {\xv}; \mathcal{P}_{\mathcal{U}}(\xv), \mathcal{U}  \right )  \,    {\rm{d}} \mu_{\mathcal{U}_C} (\xv) \nonumber \\
	& \overset{\rm (b)}{=} \!\!\!\!  \max\limits_{\substack{  \mathcal{U}  \subset \mathcal{X}: \\ \mathcal{U}= \{ \uv_k \}_{k=0}^{1}  \\  ( \uv_k )_i = \sqrt{\eta} k\tau, ~\forall k }} \!\!\!\!\!\! \alpha \int    P_e \left (\eta, {\xv}; \mathcal{P}_{\mathcal{U}_C}(\xv), \mathcal{U}  \right )  \,    {\rm{d}} \mu_{\mathcal{U}_C} (\xv) \nonumber \\
	& \overset{\rm (c)}{=} \alpha \mathsf{h}_C(\sqrt{\eta}\tau, i,2),
\end{align}
where
$\rm (a)$ follows by dropping the integral with respect to $\mu_{\Uc_D}$;
 $\rm (b)$ is due to the fact that, almost surely, with respect to $\mu_{\mathcal{U}_C} $,
%  we have that
\begin{equation}
\mathcal{P}_{\mathcal{U}}(\xv) 
%{\color{cyan} \overset{a.s.}{ = } }
= \mathcal{P}_{\mathcal{U}_C}(\xv),
\end{equation} 
where 
\begin{equation*}
\mathcal{P}_{\mathcal{U}_C}(\xv) \!=\! \left \{p_i : p_i \!=\!   \frac{  {\rm{d}} P_{\Xm_C-\uv_{i} } }{   {\rm{d} \mu_{\mathcal{U}_C}}   } (\xv),  \uv_i \in \mathcal{U}\!=\! \{ \uv_k \}_{k=0}^{\mathsf{M}-1} \right \};
\end{equation*} 
to see this note that $P_{\Xm_D}$ and $\mu_{\Uc_D}$ are singular with respect to $\mu_{\mathcal{U}_C}$ and hence, almost surely, with respect to $\mu_{\mathcal{U}_C} $ we have
\begin{align}
\frac{  {\rm{d}} P_{\Xm-\uv_{i} } }{   {\rm{d} \mu_{\mathcal{U}}}   } (\xv)= \frac{  {\rm{d}} P_{\Xm_C-\uv_{i} } }{   {\rm{d} \mu_{\mathcal{U}_C}}   } (\xv);
\end{align} 
and in $\rm (c)$ we highlight the dependence of $\mathsf{h}(\sqrt{\eta}\tau,i,2)$ on the distribution of $\Xm_C$ by using $\mathsf{h}_C(\sqrt{\eta}\tau,i,2)$ (while we use $\mathsf{h}(\sqrt{\eta}\tau,i,2)$ for the dependence on the distribution of $\Xm$).

Now, we can leverage Lemma~\ref{lem:continuous_gaussian_h} in Appendix~\ref{app:continuous_gaussian_h}  to further lower bound $\mathsf{h}_C(\sqrt{\eta}\tau,i,2) $ in~\eqref{eq:low_noise_proof_cont} as
%as $ \mathsf{h}_c(\sqrt{\eta}\tau,i,2)$ is in regards to an absolutely continuous distribution $\Xm_C$.
%Then,~\eqref{eq:low_noise_proof_cont} is further bounded by
\begin{align}\label{eq:low_noise_proof2}
	\mathsf{h}_C(\sqrt{\eta}\tau,i,2) 
	& \!\geq \! \int \! \biggl(  \int_{\yv  \in \Rc_0(\eta)}   \! \! f_\Zm(\yv \!-\! \tau \ev_i ) f_{\Xm_C}(\xv\!+ \! \sqrt{\eta} \tau \ev_i)  {\rm  d} \yv   \nonumber \\
	&~~~~  +   \int_{\yv  \in \Rc_1(\eta)} \! f_\Zm(\yv  ) f_{\Xm_C}(\xv)  {\rm  d} \yv  \,   \biggr) {\rm d}\xv,
\end{align}
where 
\begin{align*}
	& \Rc_0(\eta)  = \left\{ \yv :   \ln \frac{f_{\Xm_C}(\xv )}{f_{\Xm_C}(\xv \!+\! \sqrt{\eta}\tau \ev_i)} +  \frac{\tau^2 }{2} >  \tau y_i \right\}, \text{ and } \nonumber \\
	& \Rc_1(\eta)  = \left\{ \yv :   \ln \frac{f_{\Xm_C}(\xv)}{f_{\Xm_C}(\xv \!+\! \sqrt{\eta}\tau \ev_i)} +  \frac{\tau^2 }{2} <  \tau y_i \right\}.
%	\Rc_1 = \{ \yv \in \RR^n : & \log\frac{f_\Xm(\xv+\sigma\vv_0)}{f_\Xm(\xv+\sigma\vv_1)} \nonumber \\
%	&  +  \frac{\|\vv_1\|_2^2 - \| \vv_0 \|_2^2}{2} < (\vv_1-\vv_0)^\mathsf{T}\yv \} .
\end{align*}
In particular,  to obtain~\eqref{eq:low_noise_proof2}, we have let $\vv_0 = \zerov$ and $\vv_1 = \tau \ev_i$ in Lemma~\ref{lem:continuous_gaussian_h} in Appendix~\ref{app:continuous_gaussian_h}, and we have used $f_\Zm$ to denote the pdf of $\Zm\sim\Nc(\zerov,I_{\mathsf{d}(\Xc)})$.

%where the inequality follows from leveraging Lemma~\ref{lem:continuous_gaussian_h} with choosing $\vv_0 = \zerov$ and $\vv_1 = \tau \ev_i$ for the maximization in $\mathsf{h}_c(\sqrt{\eta}\tau,i,2) $, and denoting by $f_\Zm$ the pdf of $\Zm\sim\Nc(\zerov,I_{\dim(\Xc)})$. Note that the regions $\Rc_0(\eta)$ and $\Rc_1(\eta)$ with $\vv_0 = \zerov$ and $\vv_1 = \tau \ev_i$ are given by

Then,  by using Fatou's lemma with~\eqref{eq:low_noise_proof2}, we arrive at
\begin{align}\label{eq:low_noise_proof4}
	\liminf_{\eta\to0} \mathsf{h}_C(\sqrt{\eta}\tau,i,2)  
	& \geq  \int  \biggl(  \int_{\yv  \in \Rc_0(0)}   \! \! f_\Zm(\yv - \tau \ev_i ) f_{\Xm_C}(\xv)  {\rm  d} \yv   \nonumber \\
	&~~~~~~  +   \int_{\yv  \in \Rc_1(0)} \! f_\Zm(\yv  ) f_{\Xm_C}(\xv)  {\rm  d} \yv  \,   \biggr) {\rm d}\xv \nonumber \\
	& =  \Pr \left( |Z_i| > \frac{\tau}{2} \right).
\end{align}
Moreover, always using Fatou's lemma, we have that
\begin{align*}
%\label{eq:low_noise_proof1}
	\liminf_{\eta\to0} \!\frac{  {\mathsf{ZZ}}(P_\Xm, P_{\Ym|\Xm},2)}{\eta} 
%	& = \sum_{i=1}^n  \int_0^\infty \frac{t}{2\sigma^2} h(t,\ev_i,2)  \, {\rm d} t \nonumber \\
	& \!\!\geq \!\! \sum_{i=1}^{\mathsf{d}(\Xc)}  \!\! \!\int_0^\infty \!\!\! \liminf_{\eta\to0} \frac{\tau}{2} \mathsf{h}(\sqrt{\eta}\tau,i,2) \ {\rm d} \tau.
\end{align*}
Using~\eqref{eq:low_noise_proof_cont} and~\eqref{eq:low_noise_proof4} inside the above expression, we obtain
\begin{align}\label{eq:low_noise_proof5}
	\liminf_{\eta\to0} \frac{ {\mathsf{ZZ}}(P_\Xm, P_{\Ym|\Xm},2)}{\eta} 
	& \! \geq \alpha \!\! \sum_{i=1}^{\mathsf{d}(\Xc)} \!\! \int_0^\infty \frac{\tau}{2}  \Pr\left(|Z_i| > \frac{\tau}{2} \right) \, {\rm d} \tau \nonumber \\
	& \! = \alpha \!\! \sum_{i=1}^{\mathsf{d}(\Xc)} \! \! \EE[ Z_i^2] = \alpha \mathsf{d}(\Xc).
\end{align}
To conclude the proof of Theorem~\ref{thm:low_noise}, we note~\cite{David2016,MMSEdim} that
\begin{align}\label{eq:low_noise_proof6}
	\lim_{\eta\to0} \frac{{\rm mmse}(\Xm|\Ym)}{\eta}
%	& = \lim_{\eta\to0} \sum_{i=1}^{\dim(\Xc)} \frac{{\rm mmse}(X_i|\Ym)}{\eta} \nonumber \\
	& = \alpha \mathsf{d}(\Xc).
\end{align}
Since ${\rm mmse}(\Xm|\Ym) \geq  \! {\mathsf{ZZ}}(P_\Xm, P_{\Ym|\Xm},\mathsf{M}) \geq  {\mathsf{ZZ}}(P_\Xm, P_{\Ym|\Xm},2)$, we have that~\eqref{eq:low_noise_proof5} and~\eqref{eq:low_noise_proof6} imply that
\begin{align}
	\liminf_{\eta\to0} \frac{ {\mathsf{ZZ}}(P_\Xm, P_{\Ym|\Xm},2)}{\eta}
	& = \alpha \mathsf{d}(\Xc),
\end{align}
which completes the proof of Theorem~\ref{thm:low_noise}.

\section{Proof of Proposition~\ref{prop:tensorization}}
\label{app:sec:tensorization}

To prove the property of tensorization, it suffices to show that $\mathsf{h}(t,i,\mathsf{M}) $ in~\eqref{eq:BZZ_Mary_vec_mmse_h} depends only on $P_{X_i}$ and $P_{Y_i|X_i}$ instead of $P_{\Xm}$ and $P_{\Ym|\Xm}$.
We start by observing that
\begin{equation}
\label{eq:tensorization_proof0}
	\mathsf{h}(t,i,\mathsf{M})  
%	& =  \max\limits_{\substack{  \mathcal{U}  \subset \mathcal{X}: \\ \mathcal{U}= \{ \uv_k \}_{k=0}^{M-1}  \\  (\uv_k )_i = kt, ~\forall k }}  \int   \left(1 - \int  \max_{l\in[0:M-1]}  {\rm d} P_{\Ym|\Xm}(\yv | \xv+\uv_l )  \frac{{\rm d} P_{\Xm-\uv_l}(\xv)}{ {\rm d}\mu_\Uc(\xv)} \right) {\rm d}\mu_\Uc(\xv) \nonumber \\
	=  \mathsf{M} - \!\!\!\!\!\! \min\limits_{\substack{  \mathcal{U}  \subset \mathcal{X}: \\ \mathcal{U}= \{ \uv_k \}_{k=0}^{\mathsf{M}-1}  \\  (\uv_k )_i = kt,~\forall k }} \int  \max_{\ell\in[0:\mathsf{M}-1]}  {\rm d} P_{\Xm,\Ym}( \xv+\uv_{\ell} , \yv ) ,
\end{equation}
where we have used Lemma~\ref{lem:Pe_different_form} in Appendix~\ref{app:lem:Pe_different_form} and the fact that $\Pr(\Hc_{\ell}) = \frac{{\rm d} P_{\Xm-\uv_l}(\xv)}{ {\rm d}\mu_\Uc(\xv)}$. 
By using the facts that $P_\Xm = \prod_{i=1}^d P_{X_i}$ and $P_{\Ym|\Xm} = \prod_{i=1}^d P_{Y_i|X_i}$ with $d = \mathsf{d}(\Xc)$, we can write the integral above as
%Due to the assumption of the product probability measure, for any $s\in[1:n]\setminus i$, the integral above is lower bounded by
\begin{align}\label{eq:tensorization_proof1}
	&  \int  \max_{\ell\in[0:\mathsf{M}-1]}  {\rm d} P_{\Xm,\Ym}( \xv+\uv_{\ell }, \yv )  \nonumber \\
	& =   \int  \max_{\ell\in[0:\mathsf{M}-1]} \Biggl\{ \prod_{j=1}^d  {\rm d} P_{X_j,Y_j}( x_j+(\uv_{\ell})_j,y_j) \Biggr\} \nonumber \\
	& \geq   \int  \max_{\ell\in[0:\mathsf{M}-1]} \Biggl\{  \prod_{j\neq s}  {\rm d} P_{X_j,Y_j}( x_j+(\uv_\ell)_j, y_j)  \Biggr\} ,
\end{align}
where the inequality follows by exchanging the $\max$ with the integral, and using the fact that the probability measure over the sample space is equal to one.
By iteratively doing this for all $j$'s except for $j=i$, we obtain
\begin{align}\label{eq:tensorization_proof2}
	& \int  \max_{\ell\in[0:\mathsf{M}-1]}  {\rm d} P_{\Xm,\Ym}( \xv+\uv_\ell , \yv )  \nonumber \\
%	& \geq  \min\limits_{\substack{  \mathcal{U}  \subset \mathcal{X}: \\ \mathcal{U}= \{ \uv_k \}_{k=0}^{M-1}  \\  \left< \ev_i, \uv_k \right> = kt, \\ k \in [0:M-1] }}  \int  \max_{l\in[0:M-1]}   {\rm d} P_{X_i,Y_i}( x_i+(\uv_l)_i,y_i)   \nonumber \\
	& \geq  \int  \max_{\ell\in[0:\mathsf{M}-1]} {\rm d} P_{X_i,Y_i}( x_i+ \ell t  ,y_i)  .
\end{align}
Since the lower bound in~\eqref{eq:tensorization_proof2} can be achieved by setting $\Uc = \{\uv_k\}_{k=0}^{\mathsf{M}-1}$ such that $\uv_k = kt \ev_i,~k\in[0:\mathsf{M}-1]$ in~\eqref{eq:tensorization_proof0}, a solution for the minimization problem in~\eqref{eq:tensorization_proof0} is given by $\Uc = \{\uv_k: \uv_k = kt \ev_i,~k\in[0:\mathsf{M}-1] \}$. Hence, we have that
\begin{align}\label{eq:tensorization_proof3}
	\mathsf{h}(t,i,\mathsf{M}) 
	& = \mathsf{M} - \int  \max_{\ell\in[0:\mathsf{M}-1]} {\rm d} P_{X_i,Y_i}( x_i+ \ell t ,y_i) ,
\end{align}
which concludes the proof of Proposition~\ref{prop:tensorization} by noting that~\eqref{eq:tensorization_proof3} depends only on $P_{X_i}$ and $P_{Y_i|X_i}$.

\section{Proof of Corollary~\ref{cor:unimodal_symmetric}}
\label{app:CorIff}
Proposition~\ref{prop:mmse=ZZB} gives the sufficient and necessary condition that, for all $x,\yv$ and $t\!>\!0$, there exist $\hat{k}_1 \!=\! \hat{k}_2$ such that
\begin{subequations}
\label{eq:IffUni}
\begin{align}
	& \hat{k}_1 \in \argmax_{k\in[0:\mathsf{M}-1]}  f_{X|\Ym}(x + kt |  \yv), \text{ and} \label{eq:snc_p2_1}  \\
	& \hat{k}_2 \in \argmin_{k\in[0:\mathsf{M}-1]}  |\EE[X | \Ym = \yv] - x - kt |.  \label{eq:snc_p2_2} 
\end{align}
\end{subequations}
We first assume that $f_{X|\Ym}$ has at least two modes denoted as $m_1$ and $m_2$.
%i.e., $f_{X|\Ym}$ is multimodal. 
Then,  there exists at least one mode, say $m_1$, that is different from $\EE[X | \Ym = \yv]$. 
Now,  let $x= m_1 - \varepsilon$ and $t = \frac{2\varepsilon}{M-1}$ where $\varepsilon$ is such that $f_{X|\Ym}$ is concave in $[m_1-\varepsilon,m_1+\varepsilon]$ and  $0 < \varepsilon < |\EE[X|\Ym=\yv] - m_1|$.  This choice of $x$ and $t$ implies that $\hat{k}_1 \notin \{0, \mathsf{M}-1\}$ but $\hat{k}_2 \in \{0, \mathsf{M}-1\}$. Thus, $\hat{k}_1$ and $\hat{k}_2$ are always different, which implies that the multimodal assumption does not satisfy the condition in Proposition~\ref{prop:mmse=ZZB}. Hence, the pdf $f_{X|\Ym}$ has to be unimodal. Moreover, by using a similar argument as above, it is not difficult to show that for a unimodal pdf $f_{X|\Ym}$ with mode $m$, we need $m = \EE[X | \Ym = \yv]$ to satisfy the condition in Proposition~\ref{prop:mmse=ZZB}. 

Now, let $f_{X|\Ym}$ be unimodal and asymmetric with respect to its mode $m$.
For such a pdf, we can find $\varepsilon_0>0$ and $\varepsilon_1>0$ such that $f_{X|\Ym}(m-\varepsilon_0|\yv) = f_{X|\Ym}(m+\varepsilon_1|\yv)$. Consider the following choice for $x$ and $t>0$,
\begin{equation}
x  = m-\varepsilon_0+(-1)^{\ell}\delta~ \text{  and  }~
t = \varepsilon_0 + \varepsilon_1,
\end{equation}
where $\ell = \argmax_{i \in \{0,1\}} \varepsilon_i$ and $\delta>0$ is a small enough number. With this choice, we obtain that $\hat{k}_1 = \ell$ and $\hat{k}_2 = 1-\ell$ and hence, $\hat{k}_1\neq \hat{k}_2$, which implies that the asymmetric assumption does not satisfy the condition in Proposition~\ref{prop:mmse=ZZB}.  Hence, the pdf $f_{X|\Ym}$ has to be symmetric.

In summary, the above analysis shows that unimodality and symmetry are necessary conditions. The proof of Corollary~\ref{cor:unimodal_symmetric} is concluded by noting that~\cite{bell1995performance} showed that these are sufficient.

\section{Ancillary Lemmas}
\label{app:Lemmas}
\subsection{Lemma~\ref{lem:unimodal_pdf}}
\begin{lemma}\label{lem:unimodal_pdf}
Let $X$ be a random variable with a unimodal pdf $f_X$.  Let the mode be at $x=0$ and let $a(t)$ be
%be a constant depending on $t$
such that
\begin{subequations}\label{eq:zzb_unimodal_a0}
\begin{align}
	& \lim_{x\to a(t)^-} \max \{f_{X}(x), f_{X-t}(x)\} = f_{X-t}(a(t) ),\\
	& \lim_{x\to a(t)^+} \max \{f_{X}(x), f_{X-t}(x)\} = f_X(a(t)).
\end{align}
\end{subequations}
Then,
\begin{align}
	& \int_{-\infty}^\infty \max_{k\in[0:\mathsf{M}-1]}  f_{X-kt}(x) \ {\rm d}x \nonumber \\
&= 1 + (\mathsf{M}-1) (F_X(a(t) + t) - F_X(a(t)) ) ,
\end{align}
where $F_X$ is the cdf of $X$.
\end{lemma}
%\begin{lemma}\label{lem:unimodal_pdf}
%For an unimodal PDF of $X$, denoted by $f_X$, with a mode at $0$, let $a_t$ satisfy 
%\begin{subequations}\label{eq:lem_unimodal_at}
%\begin{align}
%	& \lim_{x\to a_t^-} \max_{k\in\ZZ} f_X(x+kt) = f_X(a_t+t), \label{eq:lem_unimodal_at_a} \\
%	& \lim_{x\to a_t^+} \max_{k\in\ZZ}  f_X(x+kt) = f_X(a_t) \label{eq:lem_unimodal_at_b}.
%\end{align}
%\end{subequations}
%Then,
%\begin{align}
%	\int_{-\infty}^\infty \max_{k\in[0:M-1]} f_{X}(x+kt){\rm d} x 
%	& = 1 + (M-1) (F_X(a_t + t) - F_X(a_t) ) ,
%\end{align}
%where $F_X$ is the CDF of $X$.
%\end{lemma}
\begin{IEEEproof}
%For the continuous PDF of $X$, we use $f_{X-kt}(x) = f_X(x+kt)$.
Since $f_X(x)$ is unimodal with mode at $x=0$, then the function $f_{X-kt}(x)$ with $k \in [0:\mathsf{M}-1]$ has maximum value at $x=-kt$.
% having the maximum value at $x=0$, the function $\max_{k\in[0:M-1]} f_{X-kt}(x)$ has the maximum value at multiple points $x=-kt,~k\in[0:M-1]$, each of which corresponds to the mode of functions $f_{X-kt}(x),~k\in[0:M-1]$. 
Moreover, the unimodality of $f_X(x)$ implies that $f_{X-kt}(x)$ is non-decreasing in $x<-kt$ and non-increasing in $x>-kt$. 
Let $a_i(t)$ for $i\in[0:\mathsf{M}-2]$ be such that the following holds,
%\begin{subequations}
\begin{align*}
	& \lim_{x\to a_i(t)^-} \max \{f_{X-it}(x), f_{X-(i+1)t}(x)\} = f_{X-(i+1)t}(a_i(t)), \\
	& \text{and }\lim_{x\to a_i(t)^+} \max \{f_{X-it}(x), f_{X-(i+1)t}(x)\} = f_{X-it}(a_i (t) ).
\end{align*}
In words, $a_i(t)$ is the transition point, that is
\begin{align*}
	\max \{f_{X\!-it}(x), f_{X\!-(i + 1)t}(x)\} 
	&\! =\! \begin{cases}
		f_{X\!-(i+1)t}(x) \! & \text{if } x \!<\! a_i(t) \\
		f_{X-it}(x) \! & \text{if } x\!> \!a_i(t) 
	\end{cases}.
\end{align*}
Note that $a_i(t)$, although it might not be unique,  always exists because of the unimodality assumption.
%{\color{blue} A.D. Why does such $a_i$ exists? Can you say in words $a_i(t)$ is? } 
%%\end{subequations}
%%Similarily, let $b_j,~j\in[1:M-1]$ be the points such that 
%%\begin{subequations}\label{eq:zzb_unimodal_proof2}
%%\begin{align}
%%	& \lim_{x\to b_j^-} \max_{k\in[0:M-1]} f_X(x+kt) = f_X(b_j +jt),\\
%%	& \lim_{x\to b_j^+} \max_{k\in[0:M-1]} f_X(x+kt) = f_X(b_j +(j-1)t).
%%\end{align}
%%\end{subequations}
%{\color{cyan}
%Note that the $a_i(t)$ is the transition point for $ \max \{f_{X-it}(x), f_{X-(i+1)t}(x)\}$, which leads us to have
%\begin{align*}
%	\max \{f_{X-it}(x), f_{X-(i+1)t}(x)\} 
%	& =\! \begin{cases}
%		f_{X-(i+1)t}(x) \! & \text{if } x < a_i(t) \\
%		f_{X-it}(x) \! & \text{if } x > a_i(t) 
%	\end{cases}.
%\end{align*}
%}
%Due to the unimodality, at $x = a_i(t)$, we have that $f_{X-jt}(x) \leq \max \{f_{X-it}(x), f_{X-(i+1)t}(x)\},~\forall j\neq i,i+1$. 
Thus,  we have that
\begin{align*}
	& \max_{k\in[0:\mathsf{M}-1]} f_{X-kt}(x) \nonumber \\
	& = \!
	\begin{cases}
		f_{X-(\mathsf{M}-1)t}(x), & \text{ if } x \!\in \!(-\infty, a_{\mathsf{M}-2}(t)) \\
		f_{X-it}(x), & \text{ if } x \! \in \! (a_{i}(t), a_{i-1}(t)),i \!\in\![1:\mathsf{M}-2] \\
		f_X(x), & \text{ if } x \! \in \!(a_0(t),\infty)
	\end{cases}.
\end{align*}
Then, we have that
\begin{align*}
%\label{eq:zzb_unimodal_proof3}
	\int_{-\infty}^\infty \max_{k\in[0:\mathsf{M}-1]} f_{X}(x+kt)\ {\rm d} x 
	& = \int_{-\infty}^{a_{\mathsf{M}-2}(t)} f_{X-(\mathsf{M}-1)t}(x) \ {\rm d} x \nonumber \\
	& + \sum_{i=1}^{\mathsf{M}-2} \int_{a_{i}(t)}^{a_{i-1}(t)} f_{X-it}(x) \ {\rm d} x \nonumber \\
	&+  \int_{a_0(t)}^{\infty} f_X(x) \ {\rm d} x .
\end{align*}
Since the pdf of $X-(i+1)t$ is shifted by $t$ to the negative direction from the pdf of $X-it$ (i.e., $f_{X-(i+1)t}(x) = f_{X-it}(x+t)$), we have that $a_{i+1}(t) +t = a_i(t)$.
%Applying the change of variable $x + it = u$ and using the fact that $a_{i+1}(t) + t = a_i(t)$ for all $i\in[0:\mathsf{M}-2]$, we 
Then, we obtain
\begin{align}\label{eq:zzb_unimodal_proof4}
	& \int_{-\infty}^\infty \max_{k\in[0:\mathsf{M}-1]} f_{X-kt}(x){\rm d} x  \nonumber \\
%	& = \int_{-\infty}^{a_{0}(t) + t} f_X(u) {\rm d} u + \sum_{i=1}^{M-2} \int_{a_0(t) }^{a_0(t) + t} f_X(u) {\rm d} u  +  \int_{a_0(t)}^{\infty} f_X(u) {\rm d} u \nonumber \\
	& = 1 + (\mathsf{M}-1)(F_X(a_0(t)+t) - F_X(a_0(t))),
\end{align}
where $F_X(x)$ is the cdf of $X$. 
The integral in~\eqref{eq:zzb_unimodal_proof4} is independent of $i\in[0:\mathsf{M}-2]$ and $a_i(t)$ and hence, we can write $a_0(t) = a(t)$, which concludes the proof of Lemma~\ref{lem:unimodal_pdf}.
\end{IEEEproof}

\subsection{Lemma~\ref{lem:continuous_gaussian_h}}\label{app:continuous_gaussian_h}

\begin{lemma}\label{lem:continuous_gaussian_h}
Let $\Ym = \Xm + \Nm$ with $\Nm\sim\Nc(\zerov_n,\eta I_n)$ and let $\Xm$ be continuous. Let $f_\Zm$ be the pdf of $\Zm\sim\Nc(\zerov_n,I_n)$. Then,
\begin{align}
	&  \mathsf{h}(t,i,2)  =\!\!\!\!\!\!  \max\limits_{\substack{  \mathcal{U}  \subset \mathcal{X}: \\ \mathcal{U}= \{ \vv_k \}_{k=0}^{1}  \\  ( \vv_k)_i= \frac{kt}{\sqrt{\eta}},~\forall k }}  \, \!\!\! \!\int \!  \left(  \int_{\yv  \in \Rc_0}   \! \!f_\Zm(\yv \!-\! \vv_1 ) f_\Xm(\xv\!+\!\sqrt{\eta} \vv_1)  {\rm  d} \yv  \right. \nonumber \\
	&\qquad\qquad  \left. +   \int_{\yv  \in \Rc_1}  f_\Zm(\yv - \vv_0 ) f_\Xm(\xv+\sqrt{\eta} \vv_0)  {\rm  d} \yv  \,   \right) {\rm d}\xv, 
\end{align}
where 
\begin{align*}
	\Rc_0  \!=\! \left\{ \yv \!:  \! \ln \frac{f_\Xm(\xv \!+\! \sqrt{\eta}\vv_0)}{f_\Xm(\xv \!+\! \sqrt{\eta}\vv_1)} \!+\!  \frac{\|\vv_1\|_2^2 \!-\! \| \vv_0 \|_2^2}{2} \!>\!  (\vv_1 \!-\! \vv_0)^\mathsf{T}\yv \right\}, \nonumber \\
	\Rc_1  \!=\! \left\{ \yv \!: \!  \ln \frac{f_\Xm(\xv \!+\! \sqrt{\eta}\vv_0)}{f_\Xm(\xv \!+\! \sqrt{\eta}\vv_1)} \!+\!  \frac{\|\vv_1\|_2^2 \!-\! \| \vv_0 \|_2^2}{2} \!<\!  (\vv_1 \!-\! \vv_0)^\mathsf{T}\yv \right\}.
%	\Rc_1 = \{ \yv \in \RR^n : & \log\frac{f_\Xm(\xv+\sigma\vv_0)}{f_\Xm(\xv+\sigma\vv_1)} \nonumber \\
%	&  +  \frac{\|\vv_1\|_2^2 - \| \vv_0 \|_2^2}{2} < (\vv_1-\vv_0)^\mathsf{T}\yv \} .
\end{align*}
\end{lemma}

\begin{IEEEproof}
The error probability for the binary hypothesis testing problem is given by
\begin{align}\label{eq:Pe_Gauss_BHT_expression}
	P_e \left ( \eta, {\xv}; \mathcal{P}_\mathcal{U}(\xv), \mathcal{U}  \right ) 
	& = \EE\left[ \min_{i\in\{0,1\}} \Pr(\Hc_i|\Ym) \right] \nonumber \\
%	& = \EE\left[ \min_{i\in\{0,1\}} \frac{f_{\Ym|\Xm}(\Ym|\xv+\uv_i) }{f_\Ym(\Ym)} \Pr(\Hc_i) \right] \nonumber \\
	& \overset{\rm (a)}{=} \EE\left[ \min_{i\in\{0,1\}} \frac{f_{\Ym|\Xm}(\Ym|\xv+\uv_i) }{f_\Ym(\Ym)} \frac{q_i}{q_0+q_1 } \right] \nonumber \\
%	& = \int  \min_{i\in\{0,1\}} f_{\Ym|\Xm}(\yv|\xv+\uv_i)  \frac{p_i }{p_0 + p_1} {\rm  d} \yv  \nonumber \\
	& = \int  \min_{i\in\{0,1\}} f_\Nm(\yv - \xv - \uv_i)   \frac{q_i }{q_0 + q_1} {\rm  d} \yv \nonumber \\
	& \overset{\rm (b)}{=} \int \! \min_{i\in\{0,1\}} \!f_\Nm(\zv - \uv_i)   \frac{q_i }{q_0 + q_1} {\rm  d} \zv ,
\end{align}
where $\rm (a)$ follows from the Bayes' rule and letting $f_\Xm(\xv+\uv_i) = q_i,~i\in\{0,1\}$, which leads to
\begin{align}
\Pr(\Hc_i) = \frac{f_\Xm(\xv+\uv_i)}{f_\Xm(\xv+\uv_0)+f_\Xm(\xv+\uv_1)} = \frac{q_i}{q_0+q_1};
\end{align}
% is because of Bayes' rule and letting $f_\Xm(\xv+\uv_i) = p_i,~i\in\{0,1\}$, 
and $\rm (b)$ follows from the change of variable $\zv = \yv - \xv$.
It is not difficult to see that~\eqref{eq:Pe_Gauss_BHT_expression} can be written as
\begin{align}\label{eq:P_e_expression_lemma_proof}
	\!\!P_e \left (\eta,{\xv}; \mathcal{P}_\mathcal{U}(\xv), \mathcal{U}  \right ) 
	& = \frac{1}{q_0\!+\!q_1}  \int_{\zv  \in \Rc_0^\prime}   f_\Zm\left(\zv - \frac{\uv_1}{\sqrt{\eta}} \!\right) q_1  {\rm  d} \zv \nonumber \\
	& \! +  \frac{1}{q_0\!+\!q_1}  \int_{\zv  \in \Rc_1^\prime}  f_\Zm\left(\zv - \frac{\uv_0}{\sqrt{\eta}} \!\right) q_0  {\rm  d} \zv,
\end{align}
where $\Zm\sim\Nc(\zerov_n, I_n)$ and
\begin{align*}
	\Rc_0^\prime  = \left\{ \zv \in \RR^n :  \ln\frac{q_0}{q_1}  +  \frac{\|\uv_1\|_2^2 - \| \uv_0 \|_2^2}{2\eta} > \frac{(\uv_1-\uv_0)^\mathsf{T}\zv}{\sqrt{\eta}} \right\}, \nonumber \\
	\Rc_1^\prime = \left\{ \zv \in \RR^n :  \ln\frac{q_0}{q_1}  +  \frac{\|\uv_1\|_2^2 - \| \uv_0 \|_2^2}{2\eta} < \frac{(\uv_1-\uv_0)^\mathsf{T}\zv}{\sqrt{\eta}} \right\}.
\end{align*}
The proof of Lemma~\ref{lem:continuous_gaussian_h} is concluded by substituting~\eqref{eq:P_e_expression_lemma_proof} inside $\mathsf{h}(t,i,2)$ in~\eqref{eq:BZZ_Mary_vec_mmse_h} with the change of variable $\uv_i = \sqrt{\eta} \vv_i$ and noticing that $\mu_\Uc(\xv) = q_0+q_1$.
%Now, we can use~\eqref{eq:P_e_expression_lemma_proof} inside in~\eqref{eq:BZZ_Mary_vec_mmse_h}
%Using change of variable $\uv_i = \sqrt{\eta} \vv_i$ in~\eqref{eq:Pe_Gauss_BHT_expression_h} and $\mu_\Uc(\xv) = p_0+p_1$, we concludes the proof of Lemma~\ref{lem:continuous_gaussian_h}.
%\begin{align}
%	\! \mathsf{h}(t,i,2)  
%%	& = \!\! \max\limits_{\substack{  \mathcal{U}  \subset \mathcal{X}: \\ \mathcal{U}= \{ \vv_k \}_{k=0}^{1}  \\  (\vv_k )_i = \frac{kt}{\sigma}, ~\forall k }}  \, \int    P_e \left ({\xv}; \mathcal{P}_\mathcal{U}(\xv), \mathcal{U}  \right )  \,    {\rm{d}} \mu_{\mathcal{U}} (\xv)    \nonumber \\
%	& =\!\! \max\limits_{\substack{  \mathcal{U}  \subset \mathcal{X}: \\ \mathcal{U}= \{ \vv_k \}_{k=0}^{1}  \\  (\vv_k)_i= \frac{kt}{\sigma}, ~\forall k }}  \, \int \left( \int_{\zv  \in \Rc_0^\prime}   f_\Zm\left(\zv - \vv_1 \right) f_{\Xm}(\xv + \sigma \vv_1)  {\rm  d} \zv \right. \nonumber \\
%	&~~~ + \left.   \int_{\zv  \in \Rc_1^\prime}  f_\Zm\left(\zv - \vv_0 \right) f_{\Xm}(\xv + \sigma \vv_0)  {\rm  d} \zv \right)  {\rm d} \xv,
%\end{align}
%where
%\begin{align}
%	\Rc_0^\prime  = \biggl\{ \zv \in \RR^n :  & \log\frac{f_\Xm(\xv+\sigma\vv_0)}{f_\Xm(\xv+\sigma\vv_1)}   \nonumber \\
%	&  +  \frac{\|\uv_1\|_2^2 - \| \uv_0 \|_2^2}{2} > (\uv_1-\uv_0)^\mathsf{T}\zv \biggr\}, \nonumber \\
%	\Rc_1^\prime = \biggl\{ \zv \in \RR^n : & \log\frac{f_\Xm(\xv+\sigma\vv_0)}{f_\Xm(\xv+\sigma\vv_1)}  \nonumber \\
%	&  +  \frac{\|\uv_1\|_2^2 - \| \uv_0 \|_2^2}{2}  < (\uv_1-\uv_0)^\mathsf{T}\zv \biggr\}.
%\end{align}
%This concludes the proof of Lemma~\ref{lem:continuous_gaussian_h}.
\end{IEEEproof}

\subsection{Lemma~\ref{lem:Pe_different_form}}
\label{app:lem:Pe_different_form}

\begin{lemma}\label{lem:Pe_different_form}
The minimum probability of error for the $\mathsf{M}$-ary hypothesis testing problem in Definition~\ref{def:MHT} can be written as
\begin{align}
	P_e \left ({\xv}; \mathcal{P}_\mathcal{U}(\xv), \mathcal{U}  \right )
	& = 1 \!-\! \int \!\! \max_{i\in[0:\mathsf{M}-1]}  \!\! {\rm d} P_{\Ym|\Xm}(\yv | \xv+\uv_i ) \ p_i.
\end{align}

\end{lemma}

\begin{IEEEproof}
The minimum error probability is attained when the MAP decision rule is used~\cite{Kay1998}. Then, for an $\mathsf{M}$-ary hypothesis testing problem as in Definition~\ref{def:MHT},
% of $\Hc_i,~i\in[0:\mathsf{M}-1]$ with observation $\Ym$, 
the minimum error probability is given by~\cite{sason2017arimoto},
\begin{align*}
%\label{eq:pe_different_form}
	P_e \left ({\xv}; \mathcal{P}_\mathcal{U}(\xv), \mathcal{U}  \right )
	& = 1 - \int  {\rm d} P_\Ym(\yv) \max_{i\in[0:\mathsf{M}-1]} \Pr(\Hc_i|\Ym=\yv) .
\end{align*}
%Recall that the hypotheses are defined in Definition~\ref{def:MHT} as $\Hc_i: \Ym \sim P_{\Ym|\Xm}(\yv|\xv+\uv_i)$ with $\Pr(\Hc_i) = \frac{{\rm d} P_{\Xm-\uv_i}(\xv)}{\sum_{k=0}^{M-1} {\rm d} P_{\Xm-\uv_k}(\xv)}$. 
Using the Bayes' rule, we obtain
\begin{align*}
	P_e \left ({\xv}; \mathcal{P}_\mathcal{U}(\xv), \mathcal{U}  \right )
	& = 1 - \int  {\rm d} P_\Ym(\yv) \max_{i\in[0:\mathsf{M}-1]} \Pr(\Hc_i|\Ym=\yv)  \nonumber \\
	& = 1 - \int  \max_{i\in[0:\mathsf{M}-1]}  {\rm d} P_{\Ym|\Hc_i}(\yv | \Hc_i ) \Pr(\Hc_i)  \nonumber \\
	& = 1 - \int  \max_{i\in[0:\mathsf{M}-1]}  {\rm d} P_{\Ym|\Xm}(\yv | \xv+\uv_i )\  p_i ,
\end{align*}
which concludes the proof of Lemma~\ref{lem:Pe_different_form}.
\end{IEEEproof}

\subsection{Lemma~\ref{lem:discrete_zero_lemma}}\label{app:discrete_zero_lemma}
\begin{lemma}\label{lem:discrete_zero_lemma}
%For any discrete $\Xm$ with its sample space $\Xc$, the following set is countable:
Assume that $\Xc$ is countable. Then, it holds that
\begin{align}
	\Bc = \Biggl\{t\geq0 : 
	& \bigcup_{\substack{\Ic\subseteq[0:\mathsf{M}-1]: \\ |\Ic| \geq 2}}  \bigcap_{k\in\Ic} \{\Xc - \uv_k\} \neq \varnothing, \uv_k\in\Xc, \nonumber \\
	& (\uv_k)_i = kt, ~k\in[0:\mathsf{M}-1] \Biggr \}
\end{align}
is countable.
\end{lemma}

\begin{IEEEproof}
Let $(a,b) \in [0:\mathsf{M}-1]^2$, $a\neq b$ and define
\begin{align}\label{eq:discrete_zero_lemma_proof1}
	\Bc_{a,b} 
	 = \{  t\geq 0 : & \{\Xc - \uv_a\} \cap  \{\Xc - \uv_b\} \neq \varnothing, \uv_a\in\Xc,\nonumber \\
	&  \uv_b\in\Xc, ( \uv_a)_i  = at, ( \uv_b )_i = bt  \}.
\end{align}
Now, observe that
\begin{align}
	\Bc  
%	& = \left\{t\geq0 : \bigcup_{\substack{\Ic\subseteq[0:M-1] \\ |\Ic| \geq 2}}  \bigcap_{k\in\Ic} \{\Xc - \uv_k\} \neq \varnothing, \uv_k\in\Xc, \left<\ev_i, \uv_k\right> = kt, k\in[0:M-1] \right \} \notag\\
%	& = \biggl\{t\geq0 : \bigcup_{\substack{\Ic\subseteq[0:M-1] \\ |\Ic| = 2}}  \bigcap_{k\in\Ic} \{\Xc - \uv_k\} \neq \varnothing, \nonumber \\
%	& \qquad\qquad\qquad \uv_k\in\Xc, ( \uv_k)_i = kt, ~k\in[0:M-1] \biggr \} \notag\\
	& = \bigcup_{\substack{(a,b)\in[0:\mathsf{M}-1]^2 \\ a\neq b}} \Bc_{a,b}.
	\label{eq:discrete_zero_lemma_proof2}
\end{align}
Since $\Xc$ is countable by assumption, then without loss of generality\footnote{When $\Xc$ is finite, the proof follows by replacing $\NN$ with $[1:|\Xc|]$.} we can assume that $\Xc = \{\xv_w : w\in\NN\}$. 
%It is not difficult to see that
%\begin{align}\label{eq:discrete_zero_lemma_proof2}
%	\Bc = \bigcup_{\substack{(a,b)\in[0:M-1]^2 \\ a\neq b}} \Bc_{a,b}.
%\end{align}
%Due to the discreteness of $\Xm$ we have that $\Xc$ is countable, and so are $\Xc - \uv_a$ and $\Xc - \uv_b$.
%Without loss of generality, assume that $\Xc$ is countably infinite\footnote{Finite $\Xc$ is a direct consequence of countably infinite $\Xc$. Specifically, by replacing $\NN$ to $|\Xc|$ in the proof we can finalize the proof even for finite $\Xc$.}. By indexing the elements in $\Xc$ we write $\Xc = \{\xv_i : i\in\NN\}$. 
Now note that, if $t\in\Bc_{a,b}$, then we have that $\xv_w - \uv_a = \xv_z - \uv_b$ for some $(w,z)\in \NN^2$. 
Thus, $\Bc_{a,b}$ in~\eqref{eq:discrete_zero_lemma_proof1} can be written as
\begin{align}\label{eq:discrete_zero_lemma_proof3}
	\Bc_{a,b} 
	& = \{  t\geq 0 : \exists (w,z)\in\NN^2, \xv_w - \xv_z = \uv_a - \uv_b,\uv_a\in\Xc,  \nonumber \\
	& \qquad\qquad\quad \uv_b\in\Xc, (\uv_a )_i= at, (\uv_b )_i = bt  \},
\end{align}
and
\begin{align}
	\Bc_{a,b} 
	& \subseteq \{  t\geq 0 : \exists (w,z)\in\NN^2, \xv_w - \xv_z = \uv_a - \uv_b, \uv_a\in\Xc, \notag \\
	& \qquad\qquad\quad \uv_b\in\Xc, ( \uv_a)_i - ( \uv_b )_i = (a-b)t \} \notag \\
%	& =  \{  t\geq 0 : \exists (i,j)\in\NN^2, \xv_i - \xv_j = \uv_a - \uv_b, \uv_a\in\Xc, \notag \\
%	& \qquad\qquad\quad  \uv_b\in\Xc, ( \uv_a)_i - ( \uv_b)_i = (a-b)t \} \notag \\
	& \subseteq \left\{  t\geq 0 : \exists (w,z)\in\NN^2, (\xv_w - \xv_z)_i = (a-b)t  \right\} \nonumber \\
	& := \overline{\Bc_{a,b} }.
	\label{eq:discrete_zero_lemma_proof4}
\end{align}
Now, note that
\begin{equation}
\overline{\Bc_{a,b} } \subseteq \left\{ \frac{ (\xv_w - \xv_z )_i}{a-b} : (w,z)\in\NN^2 \right\},
\end{equation}
which implies that 
$\overline{\Bc_{a,b} }$ is countable and hence, from~\eqref{eq:discrete_zero_lemma_proof4} we have that ${\Bc_{a,b} }$ is countable.
Since the union of countably many countable sets is still countable~\cite{folland1999real}, $\mathcal{B}$ in~\eqref{eq:discrete_zero_lemma_proof2} is countable, which concludes the proof of Lemma~\ref{lem:discrete_zero_lemma}.
\end{IEEEproof}

\section{Proof of~\eqref{eq:Bern}}\label{app:ex_discrete_ber}

From Theorem~\ref{thm:zzb_sig_inf_M}, it follows that
\begin{align}\label{eq:ex_ber_proof1}
	\mathsf{H}(P_X,t,1,  2)
	& =  \sum_{x\in\overline{\Xc}}  \max_{j\in\{0,1\}} p_{X }(x+jt)  ,
\end{align}
where $\overline{\Xc} = \cup_{j\in\{0,1\}} \{\Xc - jt \} = \{0, 1,  -t, 1 - t\}$.
It is not difficult to see that~\eqref{eq:ex_ber_proof1} is given by
\begin{align}
	\mathsf{H}(P_X,t,1,  2) = \begin{cases}
		1 + \max\{p,1-p\} & \text{ if } t = 1 \\
		2 & \text{ if } t \neq 1 
	\end{cases},
\end{align}
which results in
\begin{align}
	\Vc_t\{ 2 - \mathsf{H}(P_X,t,1,  2) \}
	& = \begin{cases}
		\min\{p,1-p\} & \text{ if } t \leq 1 \\
		0 & \text{ if } t > 1 
	\end{cases}.
\end{align}
Hence, 
%$\overline{\mathsf V}(P_X,2)  = \frac{1}{4}\min\{p,1-p\}$, and this completes the proof of~\eqref{eq:Bern}.
\begin{align}
	\overline{\mathsf V}(P_X,2) 
	& =  \int_0^\infty \frac{t}{2} \Vc_t\left\{ 2 - \mathsf{H}(P_X,t,1, 2)  \right\} \ {\rm d} t  \nonumber \\
	& = \frac{1}{4}\min\{p,1-p\},
\end{align}
and this completes the proof of~\eqref{eq:Bern}.

\end{document}